\ifdefined\ONECOLUMN
\documentclass[journal, 12pt, onecolumn, draftclsnofoot, letterpaper]{IEEEtran}
\else
\documentclass[journal, a4paper, twoside]{IEEEtran}
\fi

\IEEEoverridecommandlockouts

\usepackage{amsmath,amssymb,mathrsfs,amsbsy}
\usepackage{cite}
\usepackage{graphicx,color}
\usepackage{hyperref}
\usepackage{psfrag}
\usepackage{subfigure}
\usepackage{url}
\usepackage{balance}
\usepackage{stfloats}

\usepackage{array}
\usepackage{fancyhdr}
\usepackage{float}
\usepackage{epsfig}
\usepackage[nomain,acronym,toc]{glossaries}
\usepackage{algorithm,algorithmic}
\usepackage{caption}
\usepackage{tikz,siunitx}
\usepackage{makecell}

\newtheorem{thm}{Theorem}
\newtheorem{cor}{Corollary}[thm]

\newtheorem{lem}{Lemma}

\newtheorem{prpty}{Property}

\newtheorem{defn}{Definition}

\renewcommand{\eqref}[1]{(\ref{#1})}
\definecolor{sblue}{RGB}{0,51,160}
\definecolor{seaBlue}{RGB}{0,105,148}

\ifCLASSINFOpdf
\else
\fi

\makeatother

\begin{document}
\title{\Huge	
On Chernoff Lower-Bound of Outage Threshold for Non-Central $\chi^2$-Distributed Beamforming Gain {in URLLC Systems}
}
\author{
Jinfei Wang, Yi Ma, Rahim Tafazolli, and Zhibo Pang
\thanks{
Jinfei Wang, Yi Ma, and Rahim Tafazolli are with the 5GIC and 6GIC, Institute for Communication Systems, University of Surrey, Guildford, United Kingdom, GU2 7XH, e-mail: (jinfei.wang, y.ma, r.tafazolli)@surrey.ac.uk. 
}
\thanks{
Zhibo Pang is with the Department of Automation Technology, ABB Corporate Research Sweden, Vasteras, Sweden, and the Department of Intelligent Systems, Royal Institute of Technology (KTH), Stockholm, Sweden, email: (zhibo@kth.se). ({\it Corresponding author: Yi Ma})
}
}

\markboth{IEEE Transactions on Wireless Communications}
{IEEE Transactions on Wireless Communications}

\IEEEaftertitletext{\vspace{-2.6\baselineskip}}
\maketitle

\begin{abstract}	
The cumulative distribution function (CDF) of a non-central $\chi^2$-distributed random variable (RV) is often used when measuring the outage probability of communication systems.
For {ultra-reliable low-latency communication (URLLC)}, it is important but mathematically challenging to determine the outage threshold for an extremely small outage target. 
This motivates us to investigate lower bounds of the outage threshold, and it is found that the one derived from the Chernoff inequality (named Cher-LB) is the most effective lower bound. 
This finding is associated with {three} rigorously established properties of the Cher-LB with respect to the {mean, variance, reliability requirement, and degrees of freedom} of the non-central $\chi^2$-distributed RV.
The Cher-LB is then employed to predict the beamforming gain in URLLC for both conventional multi-antenna systems {(i.e., MIMO)} under first-order Markov time-varying channel and reconfigurable intellgent surface {(RIS)} systems.
It is exhibited that, with the proposed Cher-LB, the pessimistic prediction of the beamforming gain is made sufficiently accurate for guaranteed reliability as well as the transmit-energy efficiency.
\end{abstract}


\IEEEpeerreviewmaketitle

\section{Introduction}\label{secI}
{
\IEEEPARstart{T}{he} ultra-reliable low-latency communication (URLLC) service requires the transmitter to ensure extremely-low error probability even for a single transmission \cite{8663456,8472907}.
This elicits a fundamental problem that the transmitter must make sufficiently accurate prediction of the beamforming gain (or channel gain) to ensure the single-shot reliability when the channel knowledge at the transmitter (i.e., CSI-T) is imperfect. 
This technical requirement has been confirmed by telecommunication stakeholders such as Nokia Bell Labs and so forth \cite{8660457,8723572,9114878}. 	
Notably, the predicted beamforming gain (or channel gain) is also useful for multi-carrier system design (e.g., \cite{9520122,9963683,9171342}), as it indicates the minimum power consumption to meet the reliability requirement with respect to the channel of each carrier.
This helps the formulation of optimization problems in multi-carrier systems for URLLC \cite{Wang2022PIMRC}.

In the literature, there have been several works addressing the beamforming gain prediction problem (e.g., \cite{8660712,9120745}).
It is usually assumed that the channel is Rayleigh fading, and the receiver-side beamforming (e.g., maximum-ratio combining) is adopted.
In this case, the beamforming gain is central-$\chi^2$ distributed.
Nonetheless, the potential contribution of transmitter-side beamforming for URLLC has not been well addressed in the literature.
In this paper, we aim to investigate the beamforming gain prediction when the transmitter performs beamforming with imperfect CSI-T in multi-antenna (i.e., MIMO) systems. 
The CSI-T imperfection is assumed Gaussian, which is a widely adopted assumption when addressing the imperfections arising from channel estimation \cite{Wang2022PIMRC,1427697}) or temporal delay (e.g., \cite{891214,1593619}). 
In this case, the beamforming gain is non-central $\chi^2$-distributed.
Moreover, in reconfigurable intelligent surface (RIS) systems, the beamforming gain resembles the non-central $\chi^2$-distribution with the increase of reflector units due to the central limit theorem \cite{Basar2019,Tao2020}.
This motivates us to investigate the beamforming gain predition in RIS systems 
for URLLC as well.

Mathematically, the problem at hand translates to computing the outage threshold based on the stipulated reliability requirement.
We first present our investigation into this mathematical problem to lay the foundation for our analytical study  (see Section~\ref{secI} and Section~\ref{secII}).
Then, we present our studies in both MIMO systems (see Section~\ref{secIII}) and RIS system (see Section~\ref{secRIS}).
The mathematical problem is delineated henceforth.
}

\begin{defn}\label{defn01}
Consider $K$ independent real-Gaussian random variables (RVs), $(\alpha_0, \alpha_1, ..., \alpha_{K-1})$, with $\alpha_k\sim\mathcal{N}(\mu_k,\sigma^2)$, $_{k=0,...,K-1}$, where $\mu_k$ and $\sigma^2$ stand for their means and variances, respectively.
Define a RV, $\beta$, as follows
\begin{equation}\label{eqn01}
\beta\triangleq\sum_{k=0}^{K-1}\alpha_k^2.
\end{equation}
It is understood that $\beta$ obeys the non-central $\chi^2$-distribution with the following cumulative distribution function (CDF) \cite{AP91}
\begin{IEEEeqnarray}{rl}\label{eq02}
F_\beta(x)=1-\mathcal{Q}_{\mathrm{M},\frac{K}{2}}(\mathcal{M}/\sigma,\sqrt{x}/\sigma),
\end{IEEEeqnarray}
where $\mathcal{Q}_{\mathrm{M},\frac{K}{2}}$ stands for the generalized Marcum Q-function of order $(K)/(2)$, and $\mathcal{M}$ is defined by $\mathcal{M}\triangleq\left(\sum_{k=0}^{K-1}\mu_k^2\right)^{1/2}$.

\end{defn}

We are interested in the probability when $\beta$ is smaller than a threshold $\beta_{\mathrm{T}}$, denoted by $\mathscr{P}(\beta<\beta_{\mathrm{T}})$. 
When this probability is required to be {an extremely small} value, i.e., {$\epsilon\in(0,1)$ and $\epsilon\ll1$}, can we find the closed-form of $\beta_{\mathrm{T}}$?
\footnote{Part of this work has been presented in GLOBECOM'2023, Kuala Lumpur \cite{Wang2023a}.} 
If the answer is `{\em No}', can we find a good (or tight) lower-bound of $\beta_{\mathrm{T}}$ (denoted by $\beta^\perp\leq\beta_{\mathrm{T}})$?


{
It is worth mentioning that the reverse problem of our study, i.e., evaluating $\epsilon$ based on $\beta_{\mathrm{T}}$, is a classical problem in communication.
Broadening the scope, this includes the extensive studies on the outage probability in RIS systems of various scenarios, such as in single-RIS channels \cite{9512512} and cascaded-RIS channels \cite{9800900}.
These studies includes single-stream transmission (e.g., \cite{Basar2019,Tao2020}) as well as multi-access design (e.g., \cite{9512512,9508885}).
With respect to our problem, various bounds of the Marcum Q-function have been utilized to simplify the reverse problem (e.g., the exponential-type bound \cite{Simon2000}).
Nonetheless, they are not directly applicable to our study. 
This is because the evaluation of $\beta_\mathrm{T}$ based on $\epsilon$ remains transcendental with these bounds. 
This distinction separates our study from its reverse problem.	
}

\subsection{Prior Art}

Mathematically, the problem requires to solve the following equation
\begin{equation}\label{eqn04}
\epsilon=F_\beta(\beta_{\mathrm{T}})=\int_{0}^{\beta_{\mathrm{T}}}f_\beta(x)dx,
\end{equation} 
where $f_\beta(x)$ stands for the probability density function (PDF) of $\beta$ and is given by \cite{AP91}
\begin{equation}
f_\beta(x)=\frac{\exp\left(\frac{-(x^2+\mathcal{M}^2)}{2\sigma^2}\right)(\frac{\sqrt{x}}{\mathcal{M}})^{\frac{K}{2}-1}}{2\sigma^2}\mathcal{I}_{\frac{K}{2}-1}\left(\frac{\mathcal{M}\sqrt{x}}{\sigma^2}\right),
\end{equation}
where $\mathcal{I}(\cdot)$ stands for the modified Bessel function of the first kind.
To the best of our knowledge, there is so far no closed-form solution available in the literature. 
The problem will become even more challenging for wireless engineering when $\mathcal{M}$ is a time-varying parameter in practice. 
This is because sophisticated computing methods are often not suitable for solving a latency-sensitive mathematical problem.

Setting aside that $\epsilon$ is extremely small in URLLC systems, many approximations of $\beta_{\mathrm{T}}$ have been proposed as low-complexity solutions to \eqref{eqn04}.
For the special case when $\beta$ is central $\chi^2$-distributed, a comprehensive list of approximations can be found in \cite{Johnson1995}.
When $\beta$ is non-central $\chi^2$-distributed, Abdel-Aty's first/closer approximation{s} \cite{AbdelAty1954} and Sankaran's $z1$/$z2$ approximations \cite{Sankaran1963} are notable examples.
{These approximations demonstrate high accuracy in most cases.}
{The} principle of these approximations is to approximate a scaled version of $\sqrt{\beta}$ (or $\sqrt[3]{\beta}$) as a Gaussian RV.
Then, the scaled $\sqrt{\beta_\mathrm{T}}$ (or $\sqrt[3]{\beta_\mathrm{T}}$) is easily obtained through the inverse Gaussian Q-function.
However, the accuracy of these approximations could be questionable when $\epsilon$ is extremely small (i.e., $\beta_{\mathrm{T}}$ is small), since $\beta_{\mathrm{T}}$ could be wrongly deemed to be negative when $\beta$ is approximated as Gaussian.
Later in Fig.~\ref{figCentral} and Section~\ref{secIId}, these approximations will be compared to the Cher-LB, where our concern will be substantiated.

{
Another potential solution is using regression methodologies to fit $F_\beta(x)$ \cite{10295381,montgomery2021introduction}.
However, this solution is challenged by a fundamental trade-off: achieving an accurate fit necessitates a complex fitting function, but the inverse of $F_\beta(x)$ is only easily computable when the fitting function is simple.
Later in Section~\ref{secIId}, quadratic functions will be employed for regression to keep the inverse of $F_\beta(x)$ easily available in comparison to the proposed Cher-LB.
}

{In URLLC systems}, the lower-bound of $\beta_{\mathrm{T}}$ is of greater interest when the exact solution is not available, since a pessimistic evaluation of $\beta_{\mathrm{T}}$ can {guarantee} the single-shot reliability requirement at the cost of reasonable increase of transmission power \cite{8660712,9120745}.
These studies are comprehensively reviewed as follows.

\subsubsection{Central $\chi^2$-distribution}
Consider the case when $\mathcal{M}^2=0$ or equivalently $\mu_k=0, \forall k$. 
The PDF, $f_\beta(x)$, reduces to
\begin{equation}\label{eqn05}
f_\beta(x)=\frac{x^{\frac{K}{2}-1}\exp(\frac{-x}{2\sigma^2})}{\Gamma(K/2)(2\sigma^2)^{\frac{K}{2}}},~x>0,
\end{equation}
where $\Gamma(\cdot)$ stands for the Gamma function \cite{AP91}.
This is the classical form that has been often used to study maximum-ratio combining (MRC) in independent and identically distributed (i.i.d.) Rayleigh fading channels (e.g., \cite{8660712,8673808,9531353,8329619}).
With \eqref{eqn05}, it is still mathematically challenging to solve \eqref{eqn04}.
Nevertheless, provided $\exp(\frac{-x}{2\sigma^2})<1$ ($x>0$), the following inequality holds \cite{8660712}:
\begin{equation}\label{eqn06}
f_\beta(x)<f_{\beta,\text{poly}}(x)\triangleq\frac{x^{\frac{K}{2}-1}}{\Gamma(K/2)(2\sigma^2)^{\frac{K}{2}}}.
\end{equation}
Applying \eqref{eqn06} into \eqref{eqn04} yields
\begin{equation}\label{eqn07}
\epsilon=\int_{0}^{\beta_{\mathrm{T}}}f_\beta(x)dx<\int_{0}^{\beta_{\mathrm{T}}}f_{\beta,\text{poly}}(x)dx.
\end{equation} 
Therefore, there must exists $\beta^\perp<\beta_{\mathrm{T}}$ such that
\begin{equation}\label{eqn08}
\epsilon=\int_{0}^{\beta^\perp}f_{\beta,\text{poly}}(x)dx.
\end{equation} 
In the literature (e.g., \cite{8673808}), $\beta^\perp$ which satisfying \eqref{eqn08} is called the polynomial lower-bound (Poly-LB). 
\begin{thm}[Poly-LB from \cite{8660712}]\label{thm01}
Consider a RV, $\beta$, following the central $\chi^2$-distribution specified in \eqref{eqn05}. 
Given the probability $\epsilon$, the threshold $\beta_\mathrm{T}$ must be lower-bounded by
\begin{equation}\label{eqn09}
\beta^\perp_{\text{poly}}=2\sigma^2(\epsilon \Gamma(K/2+1))^{\frac{2}{K}}.
\end{equation} 
\end{thm}

Poly-LB has merits mainly in its compact analytical form. 
However, it can be sometimes (e.g., when $K>30$) too loose due to the inequality \eqref{eqn06}. 
In \cite{9120745}, it is shown that there exists an Chernoff lower bound (Cher-LB) for \eqref{eqn04} in the context of central $\chi^2$-distribution which {maintains a more stable tightness}. 
\begin{thm}[Cher-LB from \cite{9120745}]\label{thm02}
Given the same condition as specified in {\em Theorem~\ref{thm01}}, there exists a Cher-LB (denoted by $\beta^\perp_{\text{Cher}}$) which falls in $(0,K\sigma^2)$ and can be computed through line search based on the following equation:
\begin{equation}\label{eqn10}
\epsilon=\exp\left(\frac{K}{2}-\frac{\beta^\perp_{\text{Cher}}}{2\sigma^2}\right)\left(\frac{\beta^\perp_{\text{Cher}}}{K\sigma^2}\right)^{\frac{K}{2}}.
\end{equation}
\end{thm}

Since the right-hand term of \eqref{eqn10} is a monotonically increasing function of $\beta^\perp_{\text{Cher}}$ when $\beta^\perp_{\text{Cher}}\in(0,K\sigma^2)$, the searching complexity can be very low. 
It is worth highlighting that the form \eqref{eqn10} is not straightforwardly applicable to the context of non-central $\chi^2$-distribution;
please see our discussion in Section~\ref{secII}.

\begin{figure}[t]
\centering
\includegraphics[scale=0.53]{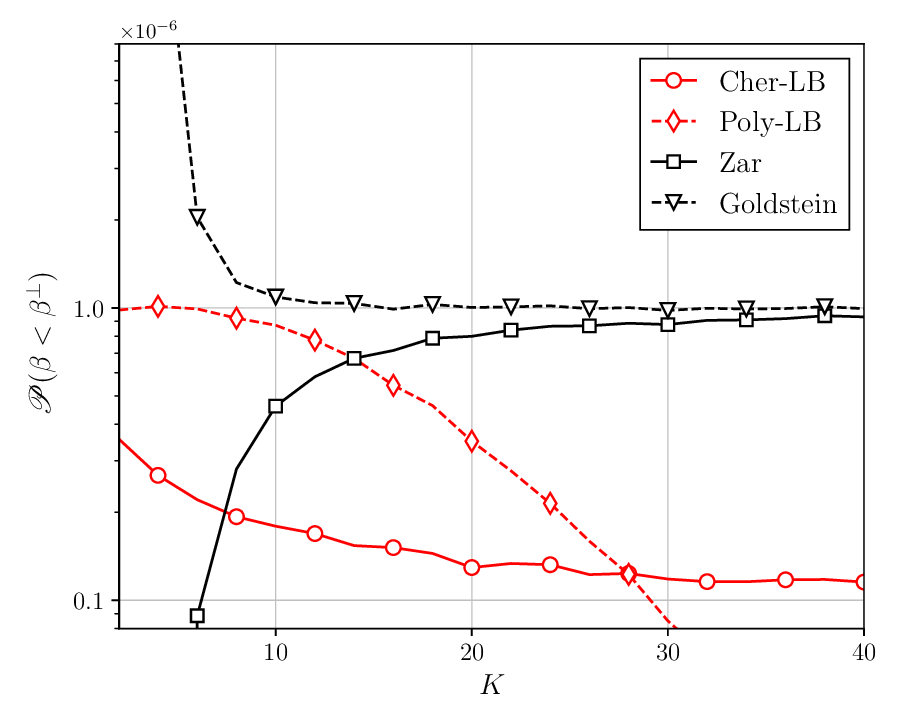}
\caption{Outage probability of various lower-bounds/approximations of $\beta_{\mathrm{T}}$ as a function of $K$ ($\sigma^2=1$ and $\epsilon=10^{-6}$).}
\vspace{-2em}
\label{figCentral}
\end{figure}

Fig.~\ref{figCentral} compares the accuracy of the aforementioned studies in terms of outage probability.
Two approximations (derived by Zar and Goldstein, respectively) are presented since they have the better accuracy than other approximations when $\epsilon$ is extremely small.
When $K<10$, both approximations show significant inaccuracy.
{In communication systems, it is very likely that $K$ is smaller than $8$ (i.e., the user has four receive-antennas or less), and using the approximations alone is therefore not suitable for URLLC.
This confirms our concern that the approximations could be inaccurate.
On the other hand, the Poly-LB shows the best accuracy when $K<15$.

}

While the Cher-LB is not as close to $\epsilon$ as the Poly-LB or the approximations, it demonstrates the most stable tightness. 
This is important for URLLC systems, where the transmitter must always predict the beamforming gain effectively, and motivates our studies in Section~\ref{secII}.
It will be shown that the Cher-LB maintains this advantage in the non-central case, and stands out as the only approach that always fulfills the outage requirement.
{Moreover, in \ref{secIIId}, it will be numerically demonstrated that the Cher-LB is sufficiently accurate for beamforming gain prediction compared to the approximations.}

\subsubsection{Non-central $\chi^2$-distribution}
For the case of $\mathcal{M}^2>0$, there are very few results about the lower-bound $\beta^\perp$ found in the literature.
{Only in \cite{8660712}, the Poly-LB was extended to the non-central case and demonstrated to keep its compact form.}
\begin{thm}\label{thm03}
Consider a RV, $\beta$, following the non-central $\chi^2$-distribution specified in \eqref{eq02}. 
The Poly-LB form of $\beta_{\mathrm{T}}$ is given by
\begin{equation}\label{eqn11}
\beta^\perp_\text{poly}=2\sigma^2(\epsilon\Gamma(K/2+1) )^{\frac{2}{K}}\exp\left(\frac{\mathcal{M}^2}{K\sigma^2}\right).
\end{equation}
\end{thm}

The Poly-LB form in \eqref{eqn11} implies $\beta^\perp_{\text{poly}}\to\infty$ as $\sigma^2\to0$.
However, it is easy to understand $\beta^\perp_{\text{poly}}\to\mathcal{M}^2$ as $\sigma^2\to0$.
This indicates that although the Poly-LB preserves its compact form, it may no longer serve as a lower-bound, as it does in the central case.
Such disadvantage arises from retaining the first Taylor expansion term of the Marcum Q-function to keep its compact form.
Retaining higher-order terms may improve $\beta^\perp_{\text{poly}}$, but it renders the closed-form solution unavailable. 
This has not been addressed in the literature, and is left for future investigation.

There are also other mathematical tools that have been commonly used to derive lower/upper bound of error probabilities. 
Those include the Jensen's inequality (e.g., \cite{7727938,8345703}), Chebyshev inequality (e.g., \cite{1650344,1542405}) as well as the aforementioned Chernoff inequality. 
Specifically, Jensen's inequality relies on the convexity of $f_\beta(x)$, and thus the lower bound is not always available. 
{As for the Chebyshev lower-bound (Cheby-LB), it is a lower-bound that measures the probability of $|\beta-\mathbb{E}(\beta)|$, where $|\cdot|$ stands for the absolute value and $\mathbb{E}(\cdot)$ for the expectation of a RV.
However, when $f_\beta(x)$ is asymmetric, the Cheby-LB can be inaccurate for the evaluation of $\beta_{\mathrm{T}}$, since $\beta_{\mathrm{T}}$ is based solely on the left tail of $f_\beta(x)$ when $\epsilon\ll1$.
Therefore, it is not a favorable lower bound for communication use-cases of interest in this paper, either.}

\subsection{Major Contribution of This Paper}

In this paper, the Cher-LB (i.e., $\beta^\perp_{\text{Cher}}$) is studied, and then employed to study MIMO {or RIS} beamforming-gain prediction in URLLC systems, demonstrating remarkable advantages in system design and optimization.
More specifically, major contributions of this work are manifolds:

\subsubsection{} 
Our study of Cher-LB starts from the generalized $\chi^2$-distribution, i.e., $\alpha_k, \forall k,$ can have different means and variances.  
It is found that $\beta^\perp_{\text{Cher}}$ can be obtained through a two-dimension (2-D) line-searching algorithm (see \textit{Theorem~\ref{thm05}} and \textbf{Algorithm~\ref{agthm1}}).
In the special case when $\beta$ conforms to the non-central $\chi^2$-distribution, \textbf{Algorithm~\ref{agthm1}} reduces to a line-searching algorithm (see \textit{Corollary~\ref{cor51}} and \textbf{Algorithm~\ref{agthm2}}).
Moreover, it is also shown that, for $\epsilon\in(0,1)$, $\beta^\perp_{\text{Cher}}$ always falls into the range of $(0,\mathbb{E}(\beta))$ with $\mathbb{E}(\beta)=\mathcal{M}^2+K\sigma^2$.
Therefore, Cher-LB could be a meaningful lower bound that can be employed to {guarantee single-shot reliability} for URLLC systems design. 

\subsubsection{} 
In communication use-cases, $\beta^\perp_{\text{Cher}}$ is expected to be as large as possible for the sake of transmit-energy saving (see Section~\ref{secIII}). 
In light of this requirement, we extensively studied the behavior of $\beta^\perp_{\text{Cher}}$ with respect to three key parameters ($\mathcal{M}^2, \sigma^2, K$).

Define the following two parameters:
\begin{subequations}
\begin{equation}\label{eqn16a}
\rho\triangleq(\mathcal{M}^2)/(\sigma^2);
\end{equation}
\begin{equation}\label{eqn16b}
\lambda\triangleq(\beta^\perp_{\text{Cher}})/(\mathcal{M}^2+K\sigma^2),
\end{equation}
\end{subequations}
where $\lambda$ is used to measure the closeness between $\beta^\perp_{\text{Cher}}$ and its maximum $(\mathcal{M}^2+K\sigma^2)$.
It is found that, when $\rho$ is fixed, $\beta^\perp_{\text{Cher}}$ scales linearly with $\mathcal{M}^2$ or $\sigma^2$ (see \textit{Property~\ref{prpty01}}).
{This motivates us to numerically study the behavior of $\beta^\perp_{\text{Cher}}$ with respect to $\rho$.}
{It is also revealed} that $\lambda$ is insensitive to the reliability requirement (e.g., the variation of outage probability constraint within the range of $(10^{-9}\sim10^{-5})$; please see \textit{Property~\ref{prpty03}}. 
In addition, \textit{Property~\ref{prpty04}} shows that, with the ratio $(\rho)/(K)$ to be fixed, $\beta^\perp_\text{Cher}$ has a diversity order of $(K)/(2)$.
These are all important results for MIMO-URLLC systems design in Section~\ref{secIII}.

\subsubsection{}
Consider the communication use-case where MIMO beamforming technology is employed to enable URLLC in the first-order Markov time-varying channel. 
Due to the channel time-variation, there is only imperfect channel knowledge available at the transmitter (i.e., imperfect CSI-T). 
The imperfectness of CSI-T introduces unknown inter-antenna/user/stream interferences, which cannot be cancelled through dirty-paper coding techniques (e.g., \cite{1056659,1603708}) particularly in not-too-large MIMO. 
In this circumstance, the system performance is dominated by interferences, where diversity combining is not a suitable technique to improve the reliability of communications \cite{1425746}. 
However, diversity combining techniques are the key to enable the extreme reliability (e.g., $10^{-9}\sim10^{-5}$ outage probability) \cite{xURLLC_nat_ele}, and thus we must consider single-stream MIMO transmission for URLLC. 
It is shown that the beamforming gain follows the non-central $\chi^2$-distribution (see \textit{Theorem~\ref{thm06}}), and the Cher-LB specified in {\em Theorem~\ref{thm05}} can be adopted to conduct power adaptation {to guarantee single-shot} reliability constraint.

\subsubsection{}
Consider the communication use-case where the RIS technology is employed to enable URLLC through passive beamforming in Rician channel.
In this case, the PDF of the beamforming gain is hardly available.
To address this issue, the beamforming gain is approximated to be non-central $\chi^2$ distributed based on the central limit theorem (CLT), and the Cher-LB is used for beamforming gain prediction (see \textit{Theorem~\ref{thm08}}).
Numerical results demonstrate that the Cher-LB effectively fulfills the single-shot reliability requirement.
With the increase of reflector units, the tightness of the Cher-LB is improving, compared to $\beta_{\mathrm{T}}$ obtained from empirical results.

\section{The Chernoff Lower Bound and Analysis}\label{secII}
\subsection{Cher-LB for Generalized $\chi^2$-Distribution}\label{secIIa}
{The mathematical foundation of our Cher-LB lies in the following well-known result for the cumulative distribution function (CDF) and its complement (i.e., the CCDF $\overline{F}_\beta(x)$):
\begin{lem}[Chernoff upper-bound from \cite{Boucheron13}]\label{lem01}
The generic Chernoff upper-bound of the CCDF/CDF for a RV (i.e., $\beta$) is given by
\begin{IEEEeqnarray}{rl}
\overline{F}_\beta(\beta_{\mathrm{T}})=\mathscr{P}(\beta\geq\beta_{\mathrm{T}})&\leq{\inf_{\nu>0}}\exp(-\nu\beta_{\mathrm{T}})\mathbb{E}(\exp(\nu\beta)),~~\\	
F_\beta(\beta_{\mathrm{T}})=\mathscr{P}(\beta\leq\beta_{\mathrm{T}})&\leq\inf_{\nu>0}\exp(\nu\beta_{\mathrm{T}})\mathbb{E}(\exp(-\nu\beta)),~~\label{eqn17}
\end{IEEEeqnarray}
where $\mathbb{E}(\exp(\nu\beta))$ is the moment generating function \cite{AP91}.
The infimum is adopted here instead of the minimum, as the right hand side of the inequalities not necessarily has an attainable minimum (e.g., see \textit{Case 1} in Appendix.~\ref{appdx2}). 
\eqref{eqn17} is of particular interest, as it serves our purpose to evaluate $F_\beta(\beta_{\mathrm{T}})$.
To facilitate our analysis, we define the following function:
\begin{equation}\label{eqn18}
s(\nu,\beta_{\mathrm{T}})\triangleq\exp(\nu\beta_{\mathrm{T}})\mathbb{E}(\exp(-\nu\beta)).
\end{equation}
Given the definition of $\beta$ in \eqref{eqn01}, $s(\nu,\beta_{\mathrm{T}})$ becomes the product of several individual terms:
\begin{equation}\label{eqn19}
	s(\nu,\beta_{\mathrm{T}})=\exp(\nu\beta_{\mathrm{T}})\prod_{k=0}^{K-1}\mathbb{E}(\exp(-\nu\alpha_k^2)).
\end{equation}
\end{lem}

}

According to \eqref{eqn07} and the discussion that follows, there exists a lower bound $\beta^\perp_\text{Cher}<\beta_{\mathrm{T}}$ which satisfies
\begin{equation}\label{eqn20}
\epsilon=\inf_{\nu>0}s(\nu,\beta^\perp_\text{Cher}),
\end{equation} 
where $\beta^\perp_\text{Cher}$ is the Cher-LB of interest. 
For the notation {simplicity}, we drop the subscript $_\text{Cher}$ in the {theoretical analysis} of Section~\ref{secIIa}-\ref{secIIc} and Section~\ref{secIIIa}-\ref{secIIIb}.

Prior to {touching} the real problem \eqref{eqn20}, it is important to acquire a closed form of $\mathbb{E}(\exp(-\nu\alpha_k^2))$ in \eqref{eqn19}.
{
It is worth noting that $\mathbb{E}(\exp(-\nu\alpha_k^2))$ is the reverse of the moment generating function $\mathbb{E}(\exp(\nu\alpha_k^2))$. Since $\mathbb{E}(\exp(\nu\alpha_k^2))$ has a closed-form expression \cite{Johnson1995}, we have}
\begin{equation}\label{eqn21}
\mathbb{E}(\exp(-\nu\alpha_k^2))=\frac{1}{\sqrt{1+2\sigma_k^2\nu}}\exp\left(-\frac{\mu_k^2\nu}{1+2\sigma_k^2\nu}\right).
\end{equation}

Then, $s(\nu,\beta^\perp)$ can be written into
\begin{equation}\label{eqn22}
s(\nu,\beta^\perp)=\frac{\psi(\nu, \beta^\perp)}{\prod_{k=0}^{K-1}(1+2\sigma_k^2\nu)^{\frac{1}{2}}},
\end{equation}
with $\psi(\nu, \beta^\perp)$ defined by
\begin{equation}\label{eqn23}
\psi(\nu, \beta^\perp)\triangleq\exp\left(\nu\beta^\perp-\sum_{k=0}^{K-1}\left(\frac{\mu_k^2\nu}{1+2\sigma_k^2\nu}\right)\right).
\end{equation}
Applying \eqref{eqn22} into \eqref{eqn20} with convexity analysis leads to the following conclusion: 
\begin{thm}\label{thm05}
Given $\epsilon\in(0,1)$, the Cher-LB, i.e., $\beta^\perp$, is a monotonically increasing function of $\epsilon$ and falls into the interval $(0,\sum_{k=0}^{K-1}(\mu_k^2+\sigma_k^2))$. 
Moreover, $\beta^\perp$ can be obtained through two-dimensional (2-D) line searching over $\beta^\perp$ and $\nu$.
\end{thm}
\begin{IEEEproof}
See Appendix~\ref{appdx2}.
\end{IEEEproof}

The 2-D line searching algorithm is provided in \textbf{Algorithm~\ref{agthm1}}, which aims to obtain an estimate of $\beta^\perp$ (i.e., $\hat{\beta}^\perp$). 
Specifically in \textbf{Algorithm~\ref{agthm1}}, $\text{sgn}$ stands for the sign-function \cite{ConcMath}. 
The variable $\nu^\star$ helps to calculate the infimum of  $s(\nu,\beta_{\mathrm{T}})$ ($\nu>0$), where
\eqref{eq26} indicates that the closed-form of $\nu^\star$ is not available, but can be obtained through line searching.

{
The complexity of \textbf{Algorithm~\ref{agthm1}} is primarily introduced by the three `while' loops.
In addition, the complexity to compute $\hat{\nu}^\star$ is different when $\nu^\star\leq\nu_\text{ini}$ or $\nu^\star>\nu_\text{ini}$.
A comprehensive breakdown of the complexity with respect to each loop is provided as follows.

\begin{itemize}
\item \textbf{Loop~1} (step~$5\sim7$): This loop aims to find the interval for searching $\nu^\star$.
When $\nu^\star\leq\nu_\text{ini}$, this loop ends immediately.
When $\nu^\star>\nu_\text{ini}$, the complexity is $\mathcal{O}(\lceil\log_{\gamma}(\nu^\star/\nu_\text{ini})\rceil)$, where $\lceil\cdot\rceil$ stands for the ceiling function \cite{ConcMath}.

\item \textbf{Loop~2} (step~$8\sim13$): This loop aims to find the numerical value of $\nu^\star$. 
When $\nu^\star\leq\nu_\text{ini}$, the searching interval is $[0,\nu_\text{ini}]$, and the complexity is $\mathcal{O}(\lceil\log_{2}(\nu_\text{ini}/\delta_\nu)\rceil)$.
When $\nu^\star>\nu_\text{ini}$, the searching interval is $[0,\nu_\text{ini}\gamma^{\lceil\log_{\gamma}(\nu^\star/\nu_\text{ini})\rceil}]$.
Notably, we have the supremum of this interval upper-bounded by a simple product:
\begin{equation}
	\nu_\text{ini}\gamma^{\lceil\log_{\gamma}(\nu^\star/\nu_\text{ini})\rceil}<\nu_\text{ini}\gamma^{\log_{\gamma}(\nu^\star/\nu_\text{ini})+1}=\gamma\nu^\star.
\end{equation}
Hence, the searching complexity is upper-bounded by $\mathcal{O}(\log_{2}(\gamma\nu^\star/\delta_\nu))$. 
We will use this bounded complexity when $\nu^\star>\nu_\text{ini}$ for its compact mathematical form.

\item \textbf{Loop~3} (step~$2\sim18$): This loop aims to find the numerical value of $\beta^\perp$. 
The complexity of line searching is $\mathcal{O}(\lceil\log_{2}(\sum_{k=0}^{K-1}(\mu_k^2+\sigma_k^2)/\delta_{\beta})\rceil)$.
\end{itemize}

The complexity to compute $\hat{\nu}^\star$ is the sum of \textbf{Loop~1}\&\textbf{2}:
\begin{IEEEeqnarray}{rl}
\mathcal{O}_{\hat{\nu}^\star}&=\mathcal{O}(\lceil\log_{2}(\nu_\text{ini}/\delta_\nu)\rceil),~\nu^\star\leq\nu_\text{ini};\label{eq25Rev}\\
\mathcal{O}_{\hat{\nu}^\star}&\approx\mathcal{O}(\lceil\log_{\gamma}(\nu^\star/\nu_\text{ini})\rceil+\log_{2}(\gamma\nu^\star/\delta_\nu)),~\nu^\star>\nu_\text{ini}.
\end{IEEEeqnarray}
Since \textbf{Loop~1}\&\textbf{2} are repeated in \textbf{Loop~3}, the overall complexity of \textbf{Algorithm~\ref{agthm1}} is the product of \textbf{Loop~3} with \textbf{Loop~1}\&\textbf{2}:
\begin{equation}
\mathcal{O}\Big(\Big\lceil\log_{2}\Big(\sum_{k=0}^{K-1}(\mu_k^2+\sigma_k^2)/\delta_{\beta}\Big)\Big\rceil\Big)\mathcal{O}_{\hat{\nu}^\star}.
\end{equation}

When $\beta^\perp\to\sum_{k=0}^{K-1}(\mu_k^2+\sigma_k^2)$, we have $\nu^\star\to0$ (see Appendix~\ref{appdx2}).
In this case, $\nu^\star$ easily falls into $(0,\nu_\text{ini})$.
When $\beta^\perp$ is small compared to $\sum_{k=0}^{K-1}(\mu_k^2+\sigma_k^2)$, it is more likely that $\nu^\star>\nu_\text{ini}$.
Nevertheless, the overall complexity is quadratic-logarithmic in both cases.
This is acceptable for real-time signal processing \cite{8815549}. 
}

\begin{algorithm}[t]
\caption{Two-Dimensional Line Searching for $\hat{\beta}^\perp$ (Generalized $\chi^2$)}
\begin{algorithmic}[1]\label{agthm1}
\renewcommand{\algorithmicrequire}{\textbf{Input:}}
\renewcommand{\algorithmicensure}{\textbf{Output:}}
\REQUIRE Outage-probability constraint $\epsilon$, $\mu_n$ and $\sigma_n^2$ for each $n$, tolerance $\delta_{\beta}$ and $\delta_\nu$, $\nu_\text{ini}$ for the initialization of $\nu$, and scaling factor $\gamma$ ($\gamma>1$) for searching $\hat{\nu}^\star$;
\ENSURE  Cher-LB $\hat{\beta}^\perp$;
\STATE \textbf{Initialization}: $\beta_\text{low}=0$, $\beta_\text{up}=\sum_{k=0}^{K-1}(\mu_k^2+\sigma_k^2)$; 
\WHILE{$\beta_\text{up}-\beta_\text{low}>\delta_{\beta}$}
\STATE $\beta_\text{mid}\leftarrow(\beta_\text{low}+\beta_\text{up})/2$;
\STATE $\nu_\text{low}\leftarrow0$, $\nu_\text{up}\leftarrow\nu_\text{ini}$;
\WHILE{$\partial s(\nu_\text{up},\beta_\text{mid})/\partial\nu_\text{up}<0$}
\STATE $\nu_\text{up}\leftarrow\gamma\nu_\text{up}$;
\ENDWHILE
\WHILE{$\nu_\text{up}-\nu_\text{low}>\delta_\nu$}
\STATE $\nu_\text{mid}\leftarrow(\nu_\text{up}+\nu_\text{low})/2$; 
\STATE $\nu_\mathrm{s}\leftarrow(\text{sgn}(\partial s(\nu_\text{mid},\beta_\text{mid})/\partial\nu_\text{mid})+1)/2$;
\STATE $\nu_\text{up}\leftarrow\nu_\mathrm{s}\nu_\text{mid}+(1-\nu_\mathrm{s})\nu_\text{up}$;
\STATE $\nu_\text{low}\leftarrow(1-\nu_\mathrm{s})\nu_\text{mid}+\nu_\mathrm{s}\nu_\text{low}$;
\ENDWHILE
\STATE $\hat{\nu}^\star\leftarrow (\nu_\text{low}+\nu_\text{up})/2$;
\STATE $\beta_\mathrm{s}\leftarrow(\text{sgn}(s(\hat{\nu}^\star,\beta_\text{mid})-\epsilon)+1)/2$;
\STATE $\beta_\text{up}\leftarrow(1-\beta_\mathrm{s})\beta_\text{mid}+\beta_\mathrm{s}\beta_\text{up}$;
\STATE $\beta_\text{low}\leftarrow\beta_\mathrm{s}\beta_\text{mid}+(1-\beta_\mathrm{s})\beta_\text{low}$;
\ENDWHILE
\STATE $\hat{\beta}^\perp\leftarrow\beta_\text{low}$;
\RETURN $\hat{\beta}^\perp$. 
\end{algorithmic} 
\end{algorithm}

\subsection{Cher-LB for Non-central $\chi^2$-Distribution}\label{secIIb}
Concerning the special case when $\beta$ is non-central $\chi^2$-distributed; as defined in \eqref{eqn01}, the algorithm used to find $\beta^\perp$ can be largely simplified. 
\begin{cor}\label{cor51}
Consider the RV, $\beta$, following the non-central $\chi^2$-distribution \eqref{eq02}.
$s(\nu,\beta^\perp)$ can be simplified into
\begin{equation}\label{eqn24}
s(\nu,\beta^\perp)=\frac{1}{(1+2\sigma^2\nu)^{\frac{K}{2}}}\exp\left(\nu\beta^\perp-\frac{\nu\mathcal{M}^2}{1+2\sigma^2\nu}\right).	
\end{equation}
There exists a closed form of the optimum $\nu$ which is expressed by
\begin{equation}\label{eqn25}
\nu^\star=\frac{K\sigma^2+\sqrt{K^2\sigma^4+4\beta^\perp\mathcal{M}^2}}{4\sigma^2\beta^\perp}-\frac{1}{2\sigma^2}.
\end{equation}
Then, \textbf{Algorithm~\ref{agthm1}} reduces to 1-D line searching in the interval of $\beta^\perp\in(0,\mathcal{M}^2+K\sigma^2)$, as shown in \textbf{Algorithm~\ref{agthm2}}.
\end{cor}
\begin{IEEEproof}
By letting $\sigma_0^2=\sigma_1^2=\cdots=\sigma_{K-1}^2=\sigma^2$, \eqref{eqn22} immediately leads to \eqref{eqn24}.
Following the proof of {\em Theorem~\ref{thm05}} and let $\partial s(\nu,\beta^\perp)/\partial \nu=0$, we obtain
\begin{equation}\label{eqn26}
\beta^\perp-\left(\frac{\mathcal{M}^2}{(1+2\sigma^2\nu)^2}+\frac{K\sigma^2}{1+2\sigma^2\nu}\right)=0.
\end{equation}
Solving \eqref{eqn26} results in \eqref{eqn25}.
With \eqref{eqn25}, step $4-13$ in \textbf{Algorithm~\ref{agthm1}} reduces to only one step, which leads to \textbf{Algorithm~\ref{agthm2}}. 
\end{IEEEproof}

It is worth noting that, when $\mathcal{M}^2=0$, \eqref{eqn24} and \eqref{eqn25} reduce to

\begin{equation}\label{eqn27a}
s(\nu,\beta^\perp)=\frac{1}{(1+2\sigma^2\nu)^{\frac{K}{2}}}\exp(\nu\beta^\perp);~\nu^\star=\frac{K}{2\beta^\perp}-\frac{1}{2\sigma^2}.
\end{equation}
{Simplifying \eqref{eqn27a}} leads to \eqref{eqn10}. 
This confirms the result discussed for the central $\chi^2$-distribution.

{
\textit{Corollary~\ref{cor51}} indicates that \textbf{Loop~1}\&\textbf{2} in \textbf{Algorithm~\ref{agthm1}} are simpified to the closed form in \eqref{eqn25}.
Hence, the complexity of \textbf{Algorithm~\ref{agthm2}} is primarily introduced by \textbf{Loop~3} compared to \textbf{Algorithm~\ref{agthm1}}.
With $\sum_{k=0}^{K-1}(\mu_k^2+\sigma_k^2)$ simplified to $\mathcal{M}^2+K\sigma^2$ as in \eqref{eqn24}, the complexity of \textbf{Algorithm~\ref{agthm2}} is given by
\begin{equation}
\mathcal{O}\Big(\Big\lceil\log_{2}\Big((\mathcal{M}^2+K\sigma^2)/\delta_{\beta}\Big)\Big\rceil\Big).
\end{equation}
This complexity is logarithmic, and is acceptable for real-time signal processing \cite{8815549}.

Besides the complexity level, it is also important to compare the complexity of the Cher-LB to the approximation approaches.
Among the approximation solutions, the Abdel-Aty's closer approximation has the highest complexity, involving three sixth-order polynomials and several lower-order ones (i.e., the complexity is roughly $\mathcal{O}(3\log_{2}(6))=\mathcal{O}(7.6)$); while the Sankaran's $z1$ approximation has the lowest complexity, involving only two quadratic polynomials (i.e., the complexity is $\mathcal{O}(2\log_{2}(2))=\mathcal{O}(2)$). 
In comparison, the complexity of \textbf{algorithm~\ref{agthm2}} is $\mathcal{O}(8)$, taking the example when $(\mathcal{M}^2+K\sigma^2)/\delta_{\beta}$ is nearly $256$. 
Hence, the complexity of \textbf{Algorithm~\ref{agthm2}} can be considered comparable to the aforementioned approximation approaches. 	
}

\begin{algorithm}[t]
\caption{Line Searching for $\hat{\beta}^\perp$ (Non-Central $\chi^2$)}
\begin{algorithmic}[1]\label{agthm2}
\renewcommand{\algorithmicrequire}{\textbf{Input:}}
\renewcommand{\algorithmicensure}{\textbf{Output:}}
\REQUIRE Outage-probability constraint $\epsilon$, $\mathcal{M}^2$, $\sigma^2$, tolerance $\delta_{\beta}$;
\ENSURE  Cher-LB $\hat{\beta}^\perp$;
\STATE \textbf{Initialization}: $\beta_\text{low}=0$, $\beta_\text{up}=\mathcal{M}^2+K\sigma^2$; 
\WHILE{$\beta_\text{up}-\beta_\text{low}>\delta_{\beta}$}
\STATE $\beta_\text{mid}\leftarrow(\beta_\text{low}+\beta_\text{up})/2$;
\STATE Calculate $\nu^\star$ based on $\beta_\text{mid}$ and \eqref{eqn25};
\STATE $\beta_\mathrm{s}\leftarrow(\text{sgn}(s(\nu^\star,\beta_\text{mid})-\epsilon)+1)/2$;
\STATE $\beta_\text{up}\leftarrow(1-\beta_\mathrm{s})\beta_\text{mid}+\beta_\mathrm{s}\beta_\text{up}$;
\STATE $\beta_\text{low}\leftarrow\beta_\mathrm{s}\beta_\text{mid}+(1-\beta_\mathrm{s})\beta_\text{low}$;
\ENDWHILE
\STATE $\hat{\beta}^\perp\leftarrow\beta_\text{low}$;
\RETURN $\hat{\beta}^\perp$. 
\end{algorithmic} 
\end{algorithm}

\subsection{Properties of Cher-LB in Non-central $\chi^2$-Distribution}\label{secIIc}
\eqref{eqn24} shows that the Cher-LB $(\beta^\perp)$ is dependent on four parameters: $\sigma^2$, $\mathcal{M}^2$, $\epsilon$ and $N$.
The aim of this subsection is therefore to study how they can impact $\beta^\perp$.
Our study starts from the scaling law of $\beta^\perp$.

\begin{prpty}[Linear scaling law]\label{prpty01}
Define a RV $\tilde{\beta}\triangleq\eta\beta$, $\eta\in\mathbb{R}^+$ with its corresponding parameters denoted by 
$\widetilde{\mathcal{M}}^2$ and $\tilde{\sigma}^2$. 
It is straightforward to obtain from \eqref{eqn01}: $\widetilde{\mathcal{M}}^2=\eta\mathcal{M}^2$ and $\tilde{\sigma}^2=\eta\sigma^2$.
Then, considering the outage probability ($\epsilon$) to be fixed, the Cher-LB of $\tilde{\beta}$ satisfies the following scaling law
\begin{equation}\label{eqn28}
\tilde{\beta}^\perp=\eta\beta^\perp.
\end{equation}
\end{prpty}
\begin{IEEEproof}
See Appendix~\ref{appdxB}.
\end{IEEEproof}

{\em Property~\ref{prpty01}} indicates that the two RVs $\tilde{\beta}$ and $\beta$ share the same $\rho$ (see the definition in \eqref{eqn16a}).
In this case, the Cher-LB satisfies the linear scaling law.
This is important for communication scenarios where the variation of $\rho$ is needed to compare the Cher-LB and the approximations when $\sigma^2\neq1$ (see Section~\ref{secIIId} and Tab.~\ref{tab02}).
Moreover, \textit{Property~\ref{prpty01}} motivates us to study the case when the ratio $\rho$ is different between two RVs, or in an equivalent sense, when $\rho$ varies for the RV $\beta$ {(see Fig.~\ref{fig3} in the later discussion)}.

Another property about the Cher-LB is the relationship between $\lambda$ and $\epsilon$.
This property is particularly interesting for mission-critical URLLC applications, where the demand for communication reliabilities can vary widely within the range of $(10^{-9}, 10^{-5})$; see \cite{8660457,8723572,9114878,3GPPTR38913}.
This motivates us to study the sensitivity of $\lambda$ with respect to a small $\epsilon$. 
As a common practice, we are interested in $\epsilon$ at the dB level, i.e.,
\begin{equation}\label{eqn31}
\varepsilon\triangleq\log_{10}(\epsilon).
\end{equation}
\begin{prpty}\label{prpty03}
$\lambda$ is a monotonically increasing function of $\varepsilon$ with the following limit for the slope
\begin{equation}\label{eqn32}
\lim\limits_{\varepsilon\to-\infty}\frac{\partial\lambda}{\partial\varepsilon}=0.
\end{equation}
\end{prpty} 
\begin{IEEEproof}
See Appendix \ref{appdxB2}.
\end{IEEEproof}

The implication of {\em Property~\ref{prpty03}} is that $\lambda$ is insensitive to the variation of $\epsilon$ when it is already small. 
This is encouraging for URLLC services that require even more stringent outage probability than $10^{-5}$ (e.g., $10^{-9}$) \cite{xURLLC_nat_ele}. 

Lastly, we are interested in the property when $\rho$ scales linearly with $K$. 
The result is  summarized as:
\begin{prpty}\label{prpty04}
Assume that $\rho$ scales linearly with $K$, i.e., $(\rho/K)=\rho_o$ (constant). 
Increasing $K$ effectively relaxes the requirement to the outage probability to $\epsilon^{\frac{2}{K}}$. 
\end{prpty}
\begin{IEEEproof}
{
The proof is rather to straightforwardly rewrite \eqref{eqn29} in Appendix~\ref{appdxB2} into
\begin{equation}\label{eqn33}
	\epsilon=\left[\frac{(1+\sqrt{1+4\lambda(\rho_o^2+\rho_o)})/(2\lambda(\rho_o+1))}{\exp\left(-\lambda(\rho_o+1)-\rho_o+\sqrt{1+4\lambda(\rho_o^2+\rho_o)}\right)}\right]^{-\frac{K}{2}}.
\end{equation}
It is understood that $\rho_o$ is a constant, and the term inside $[\cdot]$ is larger than $1$. 
Hence, increasing $K$ means to {reduce} $\epsilon$ in the order of $(K)/(2)$. 
}
\end{IEEEproof}

{In MIMO systems, the DoF is $K=2N$, where $N$ stands for the number of receive-antennas (see \textit{Theorem~\ref{thm06}}), i.e., the outage requirement is relaxed to $\epsilon^{\frac{1}{N}}$.
}

\subsection{Numerical Results of Cher-LB and Some Properties}\label{secIId}
In this subsection, numerical results are employed to demonstrate the advantages of the Cher-LB and elaborate our theoretical results.

\subsubsection*{Case Study 1}
{
This case study aims to exhibit the effectiveness of the Cher-LB in comparison to the Poly-LB (in \textit{Theorem~\ref{thm03}}), the approximations in \cite{AbdelAty1954,Sankaran1963}, as well as the regression results\footnote{We do not intend to examine the properties of the Cher-LB for the {generalized}-$\chi^2$-distribution as that is left for future work.}.

We start from examining the effectiveness by comparing $\beta^\perp$ to the numerical value of $\beta_{\mathrm{T}}$, as shown in Fig.~\ref{fig1NonCentral}. 	
For the showcase in Fig.~\ref{fig1NonCentral}, the variance of each Gaussian RV is normalized to $\sigma^2=1$. 
Moreover, the DoF is set to $K=4$ and outage requirement is $\epsilon=10^{-6}$ (related to the URLLC use cases addressed in Section~\ref{secIII}). 
Specifically, the black line stands for $\beta_{\mathrm{T}}$, the red lines for the lower-bounds, the blue lines for the approximations (including the Abdel-Aty's closer/first approximations and Sankaran's $z1$/$z2$ approximations), and the magenta lines for the regression approaches.

For the regression, we use quadratic functions here.
This is because higher-order polynomials introduce uncertainty to determine $\beta^\perp$ when inversing the fitted $F_\beta(x)$.
Moreover, $\beta_{\mathrm{T}}$ shows a quasi-linear trend when $\mathcal{M}^2$ is large as in Fig.~\ref{fig1NonCentral}.
This means exponential functions are not suitable, either.
We consider two cases: the coefficients of the quadratic function are fully flexible; or the curve is fixed to the point $(0,\beta_{\mathrm{T},0})$, where $\beta_{\mathrm{T},0}$ stands for $\beta_{\mathrm{T}}$ when $\mathcal{M}^2=0$.
The latter aims to ensure that $\beta^\perp$ is always positive.
The regression is conducted based on the least-square principle with the data of $\beta_{\mathrm{T}}$ when $\mathcal{M}^2=0,2,\cdots,200$.

The first thing to be noticed is that the Cher-LB always remains lower than $\beta_{\mathrm{T}}$, distinguishing itself as the only method to do so, as shown in Fig.~\ref{fig1NonCentral}(a).
In communication systems, this means using the Cher-LB will never overestimate the beamforming gain, and thus guarantees the single-shot reliability for URLLC.
Moreover, with the increase of $\mathcal{M}^2$, the difference the Cher-LB and $\beta_{\mathrm{T}}$ only exhibits slight increases.
This means the closeness of the Cher-LB to $\beta_{\mathrm{T}}$ (i.e., $(\beta^\perp_{\text{Cher}})/(\beta_{\mathrm{T}})$) increases with $\mathcal{M}^2$.
Overall, the Cher-LB demonstrates reasonable tightness to $\beta_{\mathrm{T}}$.
On the other hand, the Poly-LB quickly exceeds $\beta_{\mathrm{T}}$ with the increase of $\mathcal{M}^2$.
These observations coincide with our concerns to \textit{Theorem~\ref{thm03}}.

Upon examining the approximations, it is observed that the $z1$/$z2$ approximations significantly overestimate $\beta_{\mathrm{T}}$ when $\mathcal{M}^2$ is small, and the Abdel-Aty's first approximation constantly overestimate $\beta_{\mathrm{T}}$, as shown in Fig.~\ref{fig1NonCentral}(b).
However, in URLLC services, guaranteeing the expected outage probability is of utmost importance.
This means these approximations cannot guarantee the outage requirement, although the $z1$/$z2$ approximations gradually converge to $\beta_{\mathrm{T}}$ with the increase of $\mathcal{M}^2$.
The only exception here is the Abdel-Aty's closer approximation, which will be further discussed in Fig.~\ref{fig3NonCentral}.
As for the regression, both considered cases are overestimating $\beta_{\mathrm{T}}$, as shown in Fig.~\ref{fig1NonCentral}(c).
When the quadratic curve is not fixed to $(0,\beta_{\mathrm{T},0})$, $\beta^\perp$ could be negative, which is not valid as an outage threshold for $\beta$.
This is not surprising, as the fundamental trade-off mentioned in Section~\ref{secI} limits the performance of using the regression.

\begin{figure}[t!]
	\centering
	\includegraphics[scale=0.53]{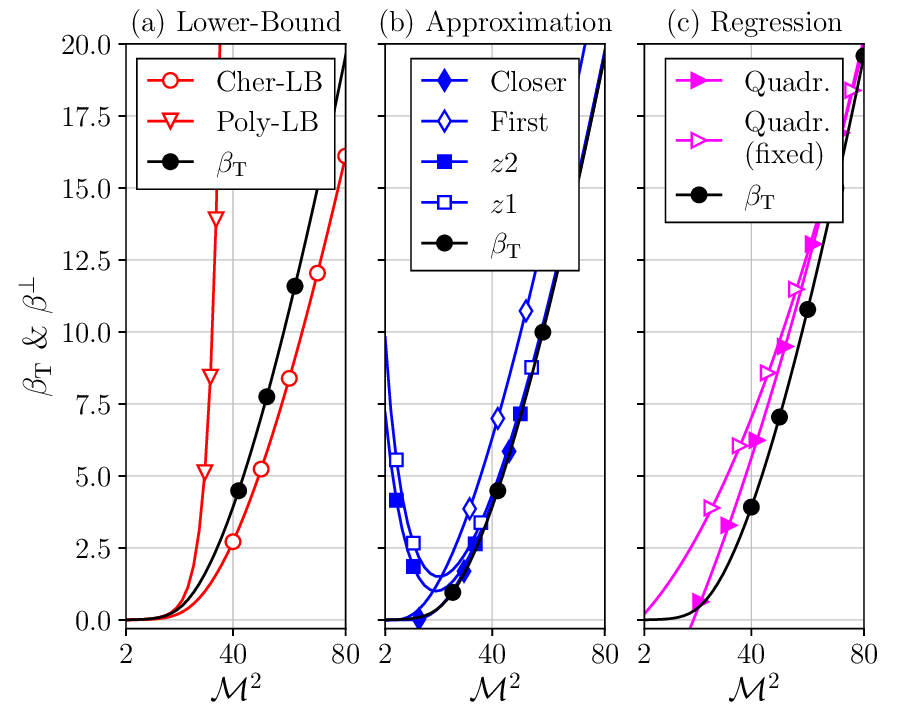}
	\caption{Lower-bounds, approximations and regressions of $\beta_{\mathrm{T}}$ as a function of $\mathcal{M}^2$ ($\sigma^2=1$, $K=4$ and $\epsilon=10^{-6}$).}
	\vspace{-1.3em}
	\label{fig1NonCentral}
\end{figure}

\begin{figure}[t]
\centering
\includegraphics[scale=0.53]{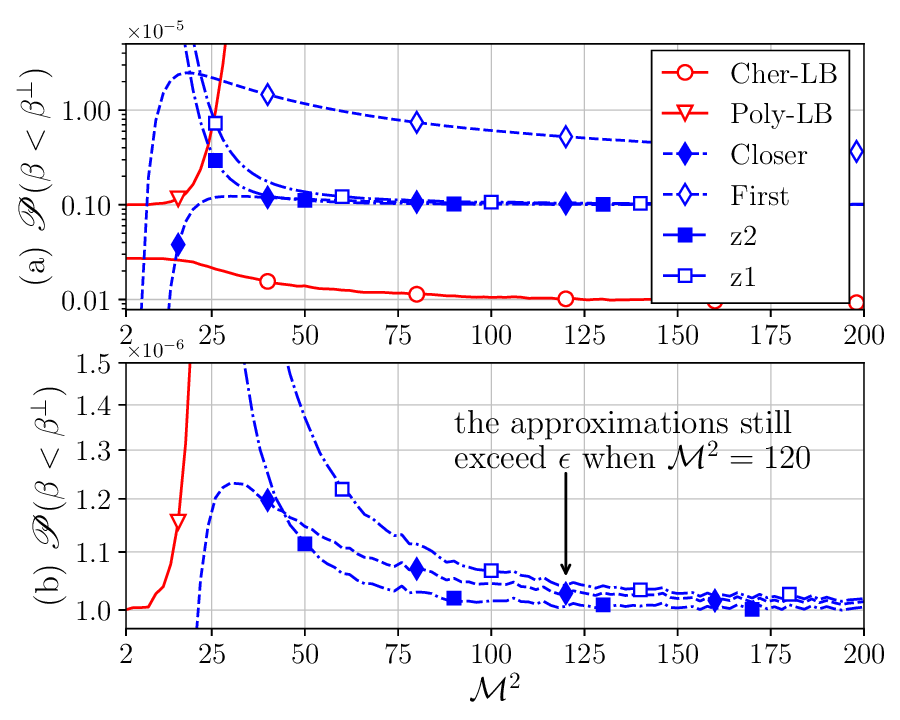}
\caption{Outage probability of lower-bounds and approximations as a function of $\mathcal{M}^2$ ($\sigma^2=1$, $K=4$ and $\epsilon=10^{-6}$).}
\vspace{-1.3em}
\label{fig3NonCentral}
\end{figure}

In order to further study the behavior of the approximations (particularly the Abdel-Aty's closer approximation), the outage probability when adopting $\beta^\perp$ is compared to the reliability requirement $\epsilon$, as shown in Fig.~\ref{fig3NonCentral}.
It is demonstrated that all approximations (including the closer approximation) have significant inaccuracy when $\mathcal{M}^2$ is small, as shown in Fig.~\ref{fig3NonCentral}(a).
Moreover, Fig.~\ref{fig3NonCentral}(b) provides a closer look, and shows that the approximations still exceeds $\epsilon$ by $2\%\sim5\%$ when $\mathcal{M}^2=120$.
Such increase of error is significant for mission-critical services.
{In Section~\ref{secIIId} Tab.~\ref{tab02}}, it will be demonstrated that there is a high probability this circumstance happens in MIMO systems.

In comparison, the Cher-LB maintains an outage probability of around $10^{-7}$, and is not sensitive to the variation of $\mathcal{M}^2$.
Hence, it is concluded that the Cher-LB is the only approach that provides an effective estimation of $\beta_{\mathrm{T}}$ for URLLC services, and has reasonable closeness to $\beta_{\mathrm{T}}$.

}

\begin{figure}[t]
\centering
\includegraphics[scale=0.53]{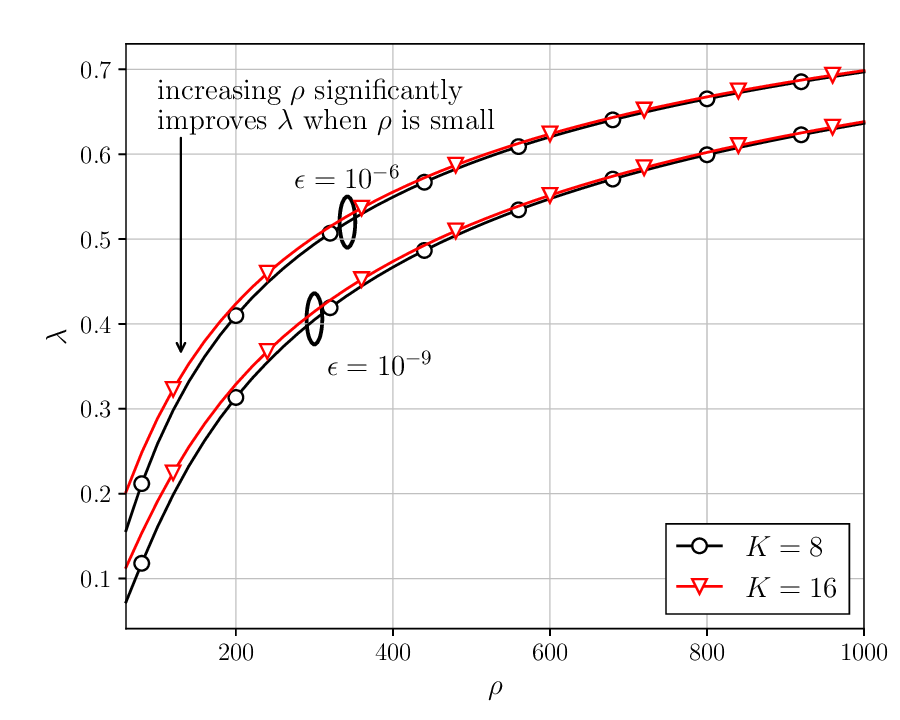}
\caption{The behavior of $\lambda$ and the approximation of $\lambda$ as $\rho$ increases when $\epsilon=10^{-6}$ and $K=8,~16,~24$, respectively.}
\vspace{-1.3em}
\label{fig3}
\end{figure}

\begin{figure}[t]
\centering
\includegraphics[scale=0.53]{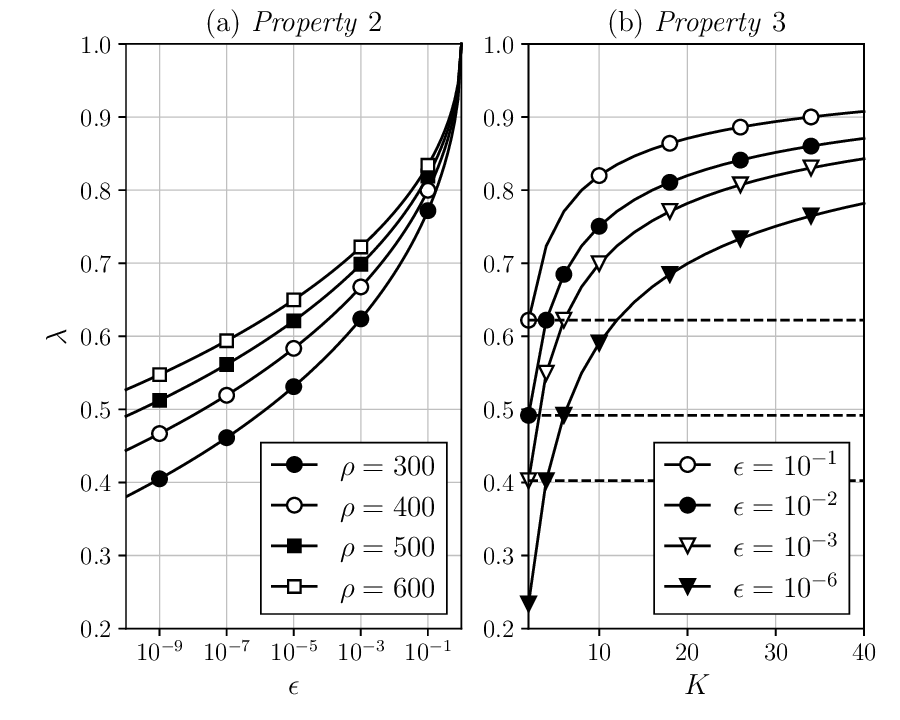}
\caption{The behavior of $\lambda$ with respect to $\epsilon$ ($K=8$) or $K$ ($\rho_o=50$) for \textit{Property~\ref{prpty03}} or \textit{Property~\ref{prpty04}}, respectively.}
\vspace{-1.3em}
\label{fig4}
\end{figure}

\subsubsection*{Case Study 2}
This case study aims to elaborate the theoretical analysis carried out in Section~\ref{secIIc}, i.e., the relationship between $\lambda$, $\epsilon$, $\rho$ and {$K$}.

{Fig. \ref{fig3} illustrates the closeness measure ($\lambda$ defined in \eqref{eqn16a}) with various configurations of $\rho$, $N$ and $\epsilon$.
It is notable that $\lambda$ grows monotonically with $\rho$. 
It grows fast when $\rho$ is small (e.g., $\rho<150$) and slow when $\rho$ is large (e.g., $\rho>150$).
This indicates that in communication systems, we need to keep $\rho$ to be not too small to avoid too much power consumption for extreme single-shot reliability.}

Fig. \ref{fig4}(a) shows the behavior of $\lambda$ when $\epsilon$ increases from $10^{-10}$ to $1$, where $\epsilon$ on the X-axis is presented in logarithm scale.
Generally, it can be observed that $\lambda$ increases monotonically with the increase of $\epsilon$.
When $\epsilon$ decreases, the slope of all curves (corresponding to different configurations of $\rho$) decreases. 
These phenomena confirm the theoretical analysis in {\em Property~\ref{prpty03}}.
More specifically, Fig.~\ref{fig4}(a) shows that, when $\epsilon$ decreases from $1$ to $10^{-5}$, $\lambda$ decreases from $1$ to somewhere around $0.6$.
When $\epsilon$ further decreases to $10^{-9}$, $\lambda$ only drops by around $0.1$.
It confirms that $\lambda$ is tolerant to the variety of reliability requirements from different mission-critical services.

Fig.~\ref{fig4}(b) shows the behavior of $\lambda$ as a function of $K$ when $\rho_o=(\rho)/(K)$ is fixed to $50$ (the conclusion holds also for other configurations).
Generally, $\lambda$ is a monotonically increasing function of {$K$}.
Concerning the case of $\epsilon=10^{-1}$, when {$K$} increases from $2$ to $12$, $\lambda$ increases by around $0.2$.
This figure increases to $0.4$ for the case of $\epsilon=10^{-6}$.
However, for both cases, further increasing {$K$} no longer induces such large improvement in $\lambda$.
Therefore, a significant gain in $\lambda$ can be obtained by increasing {$K$} when {$K$} is not large. 
Moreover, Fig.~\ref{fig4}(b) shows that increasing {$K$} is equivalent to relaxing the outage constraint to {$\epsilon^{\frac{2}{K}}$}. 
For example $\lambda$ at ({$K=2$}, $\epsilon=10^{-1}$) is equal to $\lambda$ at ({$K=4$}, $\epsilon=10^{-2}$) or $\lambda$ at ({$K=6$}, $\epsilon=10^{-3}$), as shown by the dash-lines. 
This is fully in line with the statement of diversity order in {\em Property~\ref{prpty04}}.

\section{Application of Cher-LB for MIMO-URLLC in The First-Order Markov Channel}\label{secIII}
\subsection{System Model and MIMO-URLLC Problem}\label{secIIIa}
Consider a narrowband MIMO-URLLC system, where an access point (AP) with $M$ transmit-antennas aims to communicate to users with an ultra-low outage probability (i.e., $\epsilon$).
Concerning the MIMO channel to be slowly or moderately time-varying, it can be mathematically described by the first-order Markov model (see \cite{891214,1593619})
\begin{equation}\label{eqn35}
\mathbf{H}_{t+\tau}=\mathcal{J}_0(2\pi f_\text{d}\tau)\mathbf{H}_{t}+\boldsymbol{\Omega}_{t+\tau},
\end{equation}
where $\mathbf{H}\in\mathbb{C}^{N\times M}$ stands for the MIMO channel matrix \footnote{Here, $N$ is the number of receive antennas which can be either co-located at a single user or distributed at different users. Later on, we will show that the DoF of the beamforming gain is $K=2N$ as in Section~\ref{secIIIb}.}, $t$ for the time index,  $\tau$ for the lag, $\mathcal{J}_0(\cdot)$ for the zero-order Bessel function of the first kind, $f_\text{d}$ for the maximum Doppler shift, and $\boldsymbol{\Omega}_{t+\tau}\in\mathbb{C}^{N\times M}$ for i.i.d. complex Gaussian matrix. 
Each element of $\boldsymbol{\Omega}_{t+\tau}$ follows the distribution $\mathcal{CN}(0,\sigma_\omega^2)$ with $\sigma_\omega^2=1-\mathcal{J}^2_0(2\pi f_\text{d}\tau)$. 
Since the term $(2\pi f_\text{d})$ keeps constant throughout the rest of this section, it is abbreviated in the Bessel function for the sake of notation simplicity, i.e., $\mathcal{J}_0(\tau)\triangleq\mathcal{J}_0(2\pi f_\text{d}\tau)$.

In this context, the AP (i.e., the transmitter) employs multi-antenna precoding/beamforming for the spatial multiplexing and/or spatial diversity.
To this end, the transmitter requires the CSI-T and captures it at the time slot $t$, i.e., the transmitter knows $\mathbf{H}_t$, 
and based on which the precoding/beamforming matrix $\mathbf{W}\in\mathbb{C}^{M\times L}$ is formed ($L\leq N$ stands for the number of data streams).
At the time of transmission (i.e., at the time slot $t+\tau$), \eqref{eqn35} shows that the actual CSI becomes $\mathbf{H}_{t+\tau}$, where $\mathbf{\Omega}_{t+\tau}$ is the the CSI-T uncertainty.
Therefore, the received signal at receiver(s), denoted by $\mathbf{y}_{t+\tau}$, is expressed as
\begin{IEEEeqnarray}{rl}
\mathbf{y}_{t+\tau}&=\mathbf{H}_{t+\tau}\mathbf{W}\mathbf{s}+\mathbf{v}\label{eqn36}\\
&=\mathcal{J}_0(\tau)\mathbf{H}_t\mathbf{W}\mathbf{s}+\mathbf{\Omega}_{t+\tau}\mathbf{W}\mathbf{s}+\mathbf{v},\label{eqn37}
\end{IEEEeqnarray} 
where $\mathbf{s}\in\mathbb{C}^{L\times1}$ stands for the transmitted symbol-block with $\mathbb{E}(\mathbf{s})=\mathbf{0}$ and $\mathbb{E}(\mathbf{s}\mathbf{s}^H)=E_s\mathbf{I}_{L}$ ($\mathbf{I}_{L}$: the identity matrix with the size $L$; $E_s$: the symbol energy), 
and $\mathbf{v}\in\mathbb{C}^{N\times1}$ for the additive white-Gaussian noise (AWGN) with $\mathbf{v}\sim\mathcal{CN}(\mathbf{0}, \sigma_v^2\mathbf{I}_N)$.

Consider the use of interference-rejection precoding (see \cite{1468466,Wiesel2008}), i.e., $\mathbf{H}_t\mathbf{W}=\mathbf{I}_{N}$.
To simplify our presentation, here assumes ${L}=N$ and rewrite \eqref{eqn37} into
\begin{equation}\label{eqn38}
\mathbf{y}_{t+\tau}=\mathcal{J}_0(\tau)\mathbf{s}+\mathbf{\Omega}_{t+\tau}\mathbf{W}\mathbf{s}+\mathbf{v}.
\end{equation} 
Denote $\boldsymbol{\omega}_n^T$ to be the $n^{th}$ row of $\mathbf{\Omega}_{t+\tau}$ and $\mathbf{w}_n$ to be the $n^{th}$ column of $\mathbf{W}$; the superscript $[\cdot]^T$ stands for the matrix/vector transpose.
The signal-to-interference and noise ratio (SINR) for the $n^{th}$ element of $\mathbf{y}_{t+\tau}$ is computed by
\begin{IEEEeqnarray}{ll}\label{eqn39}
\textsc{sinr}_n&=\frac{|\mathcal{J}_0(\tau)+\boldsymbol{\omega}_n^T\mathbf{w}_n|^2E_s}{\sigma_v^2+{\Upsilon_n}}~_{l,n\in\{0,...,N-1\}}\\
&\leq \frac{\mathcal{J}^2_0(\tau)E_s{+2\mathcal{J}_0(\tau)|\boldsymbol{\omega}_n^T\mathbf{w}_n|E_s}+|\boldsymbol{\omega}_n^T\mathbf{w}_n|^2E_s}{\sigma_v^2+{\Upsilon_n}},~~~~\label{eqn40}
\end{IEEEeqnarray}
where ${\Upsilon_n}=\sum_{l\neq n}|\boldsymbol{\omega}_n^T\mathbf{w}_{l}|^2E_s$ is the inter-stream interference.
Hence, current interference-rejection precoding techniques cannot eliminate the interference caused by the CSI-T uncertainty $\mathbf{\Omega}_{t+\tau}$.
The interference ${\Upsilon_n}$ is detrimental to URLLC users mainly in the sense that: 
{\em 1)} ${\Upsilon_n}$ grows $(N-1)$ times faster than the potentially contributive term $|\boldsymbol{\omega}_n^T\mathbf{w}_n|^2E_s$, and thus SINR monotonically decreases with respect to $N$; 
{\em 2)} when $\tau$ and/or $f_\text{d}$ are/is considerably large, SINR would be dominated by the interference. 
In this case, SINR reads as
\begin{equation}\label{eqn41}
\textsc{sinr}_n\approx\frac{|\mathcal{J}_0(\tau)+\boldsymbol{\omega}_n^T\mathbf{w}_n|^2}{\sum_{{l}\neq n}|\boldsymbol{\omega}_n^T\mathbf{w}_{l}|^2}.
\end{equation}
Since the SINR term \eqref{eqn41} is independent of $E_s$, it is not possible to improve the communication reliability by means of adapting the transmit-energy $E_s$. 
Then, we will have to reduce the transmission rate (i.e., increase transmission redundancy) in order to improve the reliability. 
This is however by means of trading off the spectral efficiency and more critically the latency, which is not a suitable approach for URLLC.

One might propose to use massive-MIMO because
\begin{equation}\label{eqn42}
\lim_{M\to\infty}\boldsymbol{\omega}_n^T\mathbf{w}_{l}=0, \forall {l},n,
\end{equation}
with which we have the so-called channel hardening effect
\begin{equation}\label{eqn43}
\lim_{M\to\infty}\textsc{sinr}_n=\frac{\mathcal{J}_0^2(\tau)E_s}{\sigma_v^2}.
\end{equation}
We argue that \eqref{eqn42}-\eqref{eqn43} hold not only for the interference-rejection precoding but also for the very simple matched-filter (MF) beamforming approach, i.e., $\mathbf{W}=\mathbf{H}_t^H$ ($[\cdot]^H$ denotes the matrix/vector Hermitian transpose).
Therefore, the real problem lies in the scope of not-too-large MIMO, where the interference term in \eqref{eqn39}-\eqref{eqn40} is an issue.  
In practice, interference-avoidance strategy is employed, i.e., users (or streams) are allocated on different frequencies (subchannels). 
By this means, our system of interest becomes single-stream point-to-point MIMO (${L}=1$), where the inter-stream interference is avoided. 

\subsection{Prediction of Beamforming Gain in MIMO-URLLC}\label{secIIIb}
Given the system setup of ${L}=1$ and $N>{L}$, \eqref{eqn36} becomes 
\begin{equation}\label{eqn44}
\mathbf{y}_{t+\tau}=s\mathbf{H}_{t+\tau}\mathbf{W}\mathbf{1}_N+\mathbf{v},
\end{equation}
where $\mathbf{1}_N$ denotes an $(N)\times (1)$ $1$-vector. 
It is reasonable to assume that the receiver knows $\mathbf{H}_{t+\tau}$ and $\mathbf{W}$. 
Therefore, a receiver beamforming-vector $\mathbf{u}=\mathbf{H}_{t+\tau}\mathbf{W}\mathbf{1}_N$ can be employed to enable the {MRC} as
\begin{IEEEeqnarray}{ll}\label{eqn45}
y&=\mathbf{u}^H\mathbf{y}_{t+\tau}\\
&=\|\mathbf{H}_{t+\tau}\mathbf{W}\mathbf{1}_N\|^2s+\tilde{v},\label{eqn46}
\end{IEEEeqnarray}
where $\tilde{v}$ is the corresponding AWGN after the receiver beamforming.  
The signal-to-noise ratio (SNR) is computed by
\begin{equation}\label{eqn47}
\textsc{snr}=\frac{\|\mathbf{H}_{t+\tau}\mathbf{W}\mathbf{1}_N\|^2E_s}{\sigma_v^2}.
\end{equation}
Hence, the beamforming gain of this system is given by
\begin{IEEEeqnarray}{ll}\label{eqn48}
\beta&=\|\mathbf{H}_{t+\tau}\mathbf{W}\mathbf{1}_N\|^2\\
&=\|\mathcal{J}_0(\tau)\mathbf{H}_t\mathbf{W}\mathbf{1}_N+\mathbf{\Omega}_{t+\tau}\mathbf{W}\mathbf{1}_N\|^2.\label{eqn49}
\end{IEEEeqnarray} 

\begin{thm}\label{thm06}

Let $\boldsymbol{\alpha}=\mathbf{H}_{t+\tau}\mathbf{W}\mathbf{1}_N$.
Then, $\beta$ is the sum of squares of $2N$ real-Gaussian RVs: $\beta=\sum_{n=0}^{N-1}(\Re(\alpha_n)^2+\Im(\alpha_n)^2)$.
$\Re(\alpha_n)$ and $\Im(\alpha_n)$, $_{n=0,1,\cdots N-1}$, have identical variance:
\begin{equation}\label{eqn50}
\sigma^2=\sigma_{\omega}^2\|\mathbf{W}\mathbf{1}_N\|^2/2.
\end{equation}
The mean of $\alpha_n$ is given by
\begin{equation}\label{eqn51}
\mu_n=\mathcal{J}_0(\tau)\mathbf{h}^T_{t,n}\mathbf{W}\mathbf{1}_N,
\end{equation}
where $\mathbf{h}^T_{t,n}$ is the $n^{th}$ row vector of $\mathbf{H}_{t}$.
Hence, the mean of $\Re(\alpha_n)$ and $\Im(\alpha_n)$ is given by $\Re(\mu_n)$ and $\Im(\mu_n)$, respectively.

According to {\em Definition~\ref{defn01}}, $\beta$ obeys the non-central $\chi^2$-distribution with $K=2N$ and 
\begin{equation}\label{eqn52}
\mathcal{M}^2=\mathcal{J}_0^2(\tau)\sum_{n=0}^{N-1}\|\mathbf{h}^T_{t,n}\mathbf{W}\mathbf{1}_N\|^2.
\end{equation}

\end{thm}
\begin{IEEEproof}
The proof is rather straightforward and thus omitted for the sake of spacing limit. 
\end{IEEEproof}

{\em Theorem \ref{thm06}} shows that $\beta$ is a non-central $\chi^2$-distributed RV. 
Given the reliability constraint $\epsilon$, we can use {\em Corollary~\ref{cor51}} and {\bf {Algorithm \ref{agthm2}}} to obtain the Cher-LB, $\beta^\perp$.
This helps the transmitter to know 
\begin{equation}\label{eqn53}
\mathscr{P}\Bigl(\textsc{snr}\geq\frac{\beta^\perp E_s}{\sigma_v^2}\Bigl)\geq1-\epsilon.
\end{equation}
{and scales $E_s$ accordingly to guarantee the reliability.}

The point-to-point MIMO-URLLC system presented in Section~\ref{secIIIb} assumes that the receiver knows $\mathbf{H}_{t+\tau}$.
This motivates us to study the channel hardening behavior of the beamforming gain ($\beta$) when $M\to\infty$.
Given that $\mathbf{W}$ is uncorrelated with $\mathbf{\Omega}_{t+\tau}$, we immediately have
\begin{equation}\label{eqn55}
\lim_{M\to\infty}\beta=\mathcal{J}_0^2(\tau)\|\mathbf{H}_t\mathbf{W}\mathbf{1}_N\|^2\mathop{\leq}^\text{(a)}\mathcal{J}_0^2(\tau),
\end{equation}
where the equality on {(a)} holds when $\mathbf{W}$ is designed to fulfill $\|\mathbf{H}_t\mathbf{W}\mathbf{1}_N\|^2=1$.
This result is in line with \eqref{eqn43}.
Unsurprisingly, in the presence of CSI-T uncertainty, we cannot improve the beamforming gain from the receiver side. 
{
Nevertheless, our major concern lies in not-too-large MIMOs as stated in Section~\ref{secIIIa}, where increasing the transmit/receive-antennas can still make significant improvement on the predicted beamforming gain.
}

\subsection{Numerical Results and Discussion}\label{secIIId}
In this subsection, numerical results are employed to demonstrate the performance of the Cher-LB in MIMO systems under first-order Markov channel and elaborate our theoretical results. As {discussed} in Section~\ref{secIIIa}, we consider single-stream point-to-point MIMO, where the receiver (i.e., user) only has a small number of antennas. This means our MIMO system is equivalently a massive-MIMO downlink. $\mathbf{H}_t$ is assumed to be i.i.d. Rayleigh \cite{8660712,8673808}, and MF beamforming (i.e., $\mathbf{W}=\mathbf{H}_t^H$) is adopted in appreciation of its low computational complexity and reasonable performance in massive-MIMO (e.g., \cite{1564285,1468466}). Since the beamforming gain is normalized, $\delta_{\beta}$ of {\textbf{Algorithm~\ref{agthm1}}} is reduced to $10^{-4}$. The central carrier frequency is assumed to be $3.5$ GHz (see \cite{Euro5GObservatory}). Considering the stringent latency requirement of URLLC, the time lag $\tau$ is set to be $0.5$ ms. The user velocity is assumed to be $20$ m/s {($72$ km/h)}, which is a pessimistic consideration of vehicular velocity in urban areas.

\begin{table}[t]
\center
\caption{$\mathscr{P}(\rho<120)$}
\label{tab02}
\renewcommand{\arraystretch}{1.0}
{
\begin{tabular}{ c !{\vrule width1pt} c | c | c | c | c  } 
\Xhline{1.6pt} 
$M$ & $16$ & $20$ & $24$ & $28$ & $32$  \\
\Xhline{1pt} 
$N=2$ & $72.7\%$ & $37.5\%$ & $11.8\%$ & $2.28\%$ & $0.273\%$ \\ 
\hline
$N=4$ & $55.4\%$ & $22.4\%$ & $5.51\%$ & $0.848\%$ & $0.0794\%$ \\ 
\Xhline{1.6pt} 
\end{tabular}
}
\vspace{-1.3em}
\end{table}

{\em Numerical Example 1:}
The aim of this example is to demonstrate the Cher-LB of MF beamforming gain in MIMO systems under first-order Markov channel. Monte-Carlo trials are performed to study the properties of interest.

The first thing of interest is the probability that only the Cher-LB is effective in communication systems.
In Fig.~\ref{fig1NonCentral}, this is when $\mathcal{M}^2<120$. 
Since in communication systems $\sigma^2$ is usually not unit, this probability is equivalently defined as $\mathscr{P}(\rho<120)$ based on \textit{Property~\ref{prpty01}}.
Tab.~\ref{tab02} shows $\mathscr{P}(\rho<120)$ when $N=2$ or $4$ as the transmit-antenna number $M$ increases from $16$ to $32$. 
It is shown that $\mathscr{P}(\rho<120)$ is dominant when the transmit-antenna number is not large (i.e., $M<24$).
With the increase of $M$, $\mathscr{P}(\rho<120)$ is decreasing. 
This is because increasing $M$ enhances the beamforming gain, and makes $\rho$ larger.
However, when $M=32$, $\mathscr{P}(\rho<120)$ is still significant compared to the outage requirement $\epsilon$ ($10^{-5}$ or less).
This means without the Cher-LB, the failure probability of the beamforming gain prediction will exceed $\epsilon$.
In conclusion, the Cher-LB is necessary for beamforming gain prediction when the MIMO size is not too large.
Note that since $N$ is small, the configurations in Tab.~\ref{tab02} still suit the definition of massive-MIMO \cite{1468466}.

\begin{figure}[t]
\centering
\includegraphics[scale=0.53]{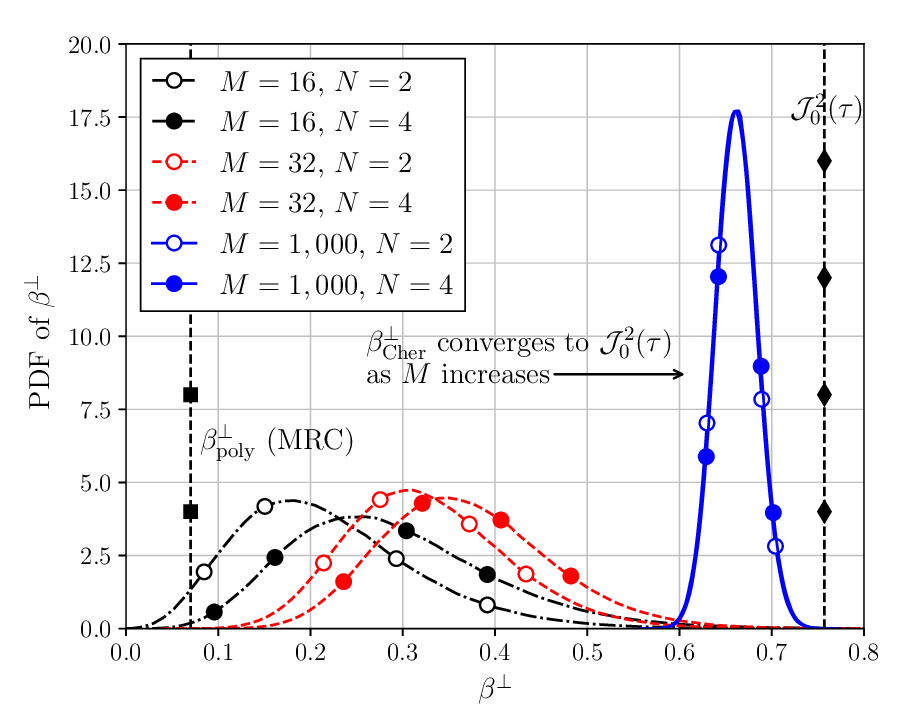}
\caption{The PDF of $\beta^\perp_\text{Cher}$ as $M$ increases from $16$ to $32$ and $1,000$ when using MF beamforming, $\epsilon=10^{-6}$, $N=2$ or $4$.}
\vspace{-1.3em}
\label{fig7}
\end{figure}

{
After showing the necessity of using the Cher-LB, we are interested in comparing the predicted beamforming gain to using MRC alone, as shown in Fig.~\ref{fig7}
We consider the case where $N=2$ (marked by circles) or $4$ (marked by triangles) \cite{3GPPTR38913}. 
For MRC, only the Poly-LB of $N=4$ (i.e., $K=8$) is presented as it shows the best performance in Fig.~\ref{figCentral}.
It is observed that using beamforming gain prediction for MF beamforming significantly outperforms using MRC alone.
When $M=16$, $\beta^\perp_{\text{Cher}}$ achieves around $0.17$, compared to $\beta^\perp_{\text{poly}}=0.07$.
Moreover, as $M$ increases to $32$, $\beta^\perp_{\text{Cher}}$ increases to around $0.3$.
This is because a larger $M$ brings more beamforming gain (i.e., higher $\rho$).
As $M$ increases to $1,000$, $\beta^\perp_\text{Cher}$ approaches the limit of $\mathcal{J}_0^2(\tau)$. 
This coincides with the discussions in Section~\ref{secIIIb}.  
In addition, increasing $N$ also improves $\beta^\perp_{\text{Cher}}$ when $M$ is not too large.
This is because $\mathbf{H}_t$ becomes quasi-orthogonal with the increase of $M$ (e.g., $M=1,000$), and increasing $N$ makes little difference in this case.
Nevertheless, when $M$ is $16$ or $32$, the improvement of increasing $N$ is significant ($50\%$ or $20\%$, respectively).
}

\begin{figure}[t]
\centering
\includegraphics[scale=0.53]{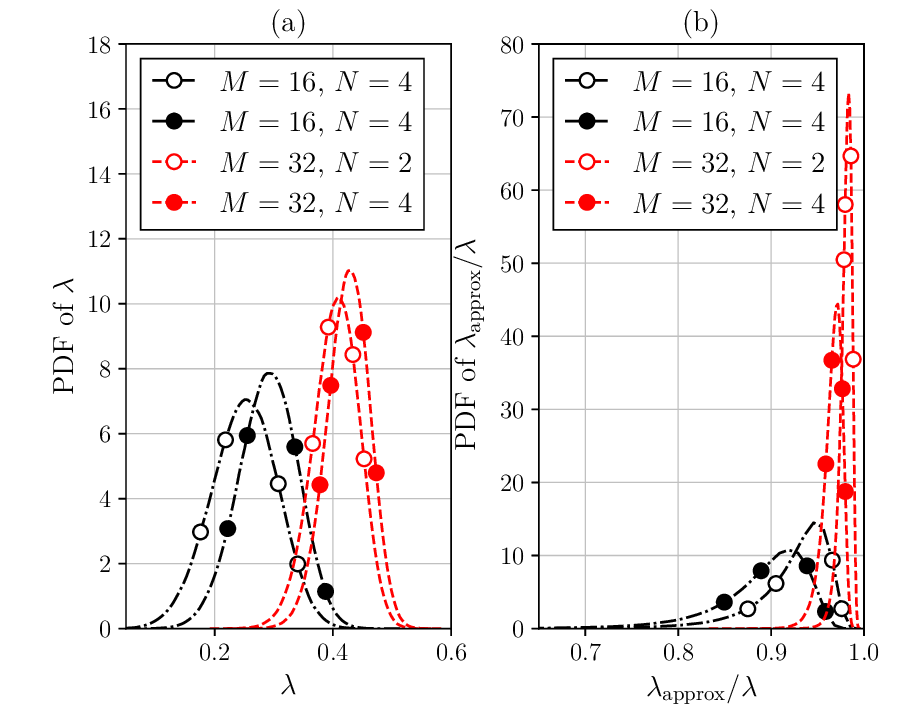}
\caption{The PDF of $\lambda$ and its approximation as $M$ increases from $16$ to $32$ when using MF beamforming, $\epsilon=10^{-6}$, $N=2$ or $4$.}
\vspace{-1.3em}
\label{fig8aa}
\end{figure}

{
The next thing of interest is the comparison of beamforming gain in URLLC systems and conventional throughput-oriented systems (i.e., $\lambda$), as shown in Fig.~\ref{fig8aa}.
It is observed that as $M$ increases from $16$ to $32$, $\lambda$ increases from around $0.25$ to around $0.4$.
This means it costs $2.5\sim5$ times of power to deliver URLLC service compared to throughput-oriented service.
Considering the strict reliability requirement, this power consumption is reasonable.}

\begin{figure}[t]
\centering
\includegraphics[scale=0.53]{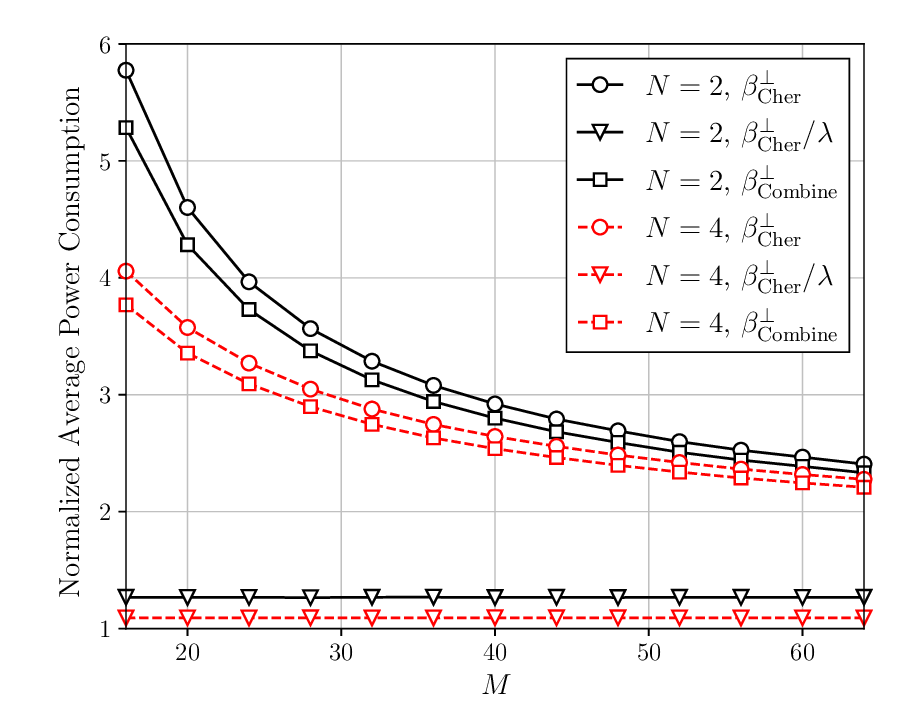}
\caption{The normalized average power consumption for power adaptation when $\epsilon=10^{-6}$, $M$ increases from $16$ to $64$.}
\vspace{-1.3em}
\label{fig8}
\end{figure}

Finally, we are interested in the average power consumption of the power adaptation based on the predicted beamforming gain, as shown in Fig.~\ref{fig8}.
With a fixed SNR target, $E_\text{s}$ linearly scales to $(1)/(\beta^\perp_\text{Cher})$.
Hence, we use $\mathbb{E}((1)/(\beta^\perp_\text{Cher}))$ as the normalized average power consumption. 
In addition to using the Cher-LB alone, we consider a combination of the Cher-LB and the $z2$ approximation.
Based on Fig.~\ref{fig3NonCentral} and \textit{Property~\ref{prpty01}}, the predicted beamforming gain is chosen to be the $z2$ approximation when $\rho\geq120$.
These approaches are compared to the throughput-oriented system, i.e., the normalized average power consumption is $\mathbb{E}((\lambda)/(\beta^\perp_{\text{Cher}}))$.

Compared to $\mathbb{E}((\lambda)/(\beta^\perp_{\text{Cher}}))$, which overall stays around $1.1\sim1.25$, $\mathbb{E}((1)/(\beta^\perp_{\text{Cher}}))$ decreases from around $5.5$ to $2.2$ as $M$ increases from $16$ to $64$.
Considering the extreme reliability requirement, this is a reasonable power consumption.
Moreover, the combination approach only shows around $0.2$ improvement compared to using the Cher-LB alone.
This is reasonable, since when $M$ is small, the probability to use the $z2$ approximation is low, while when $M$ is large, the Cher-LB itself is close to the approximation.
Overall, using the Cher-LB alone achieves reasonable average power consumption.

\section{Application of Cher-LB for RIS-Aided URLLC}\label{secRIS}
\subsection{System Model and CLT Approximation}

Besides conventional MIMO, recent studies have incorporated RIS systems to facilitate URLLC services \cite{Ren2022, Liu2021RIS, Yang2020}.
The RIS systems are expected to support URLLC services by providing supplementary {line-of-sight} links and improving the beamforming gain through phase alignment.
Although the outage probability has been a concern in the literature (e.g., \cite{Ren2022, Liu2021RIS, Yang2020}), the performance of beamforming gain prediction to guarantee single-shot outage probability in RIS-enabled URLLC systems remains to be investigated.
This motivates our investigation into the application of the Cher-LB for beamforming gain prediction in RIS systems.
To this end, the CLT is employed to approximate the beamforming gain to be non-central $\chi^2$-distributed (e.g., \cite{Yang2020,Basar2019,Abdullah2020,Tao2020}).
Numerical results corroborate the effectiveness of the Cher-LB in fulfilling the outage requirement.

Consider a RIS system with single transmit-antenna, single receive-antenna, and a RIS with $N_\mathrm{R}$ reflector units.
It is assumed that the signal only reaches the receiver with the aid of the RIS (e.g., the direct link may not exist due to blockage) \cite{Ren2022,Yang2020,Basar2019,Abdullah2020}.
The RIS shifts the phase of the signal through the reflector units. 
Then, the received signal is given by
\begin{equation}
y=\mathbf{g}^T\mathbf{\Phi}\mathbf{h}s+v,
\end{equation}
where $\mathbf{g}^T\in\mathbb{C}^{1\times N_\mathrm{R}}$ stands for the channel from the RIS to the receiver, $\mathbf{h}\in\mathbb{C}^{N_\mathrm{R}\times 1}$ for the channel from the transmitter to the RIS, and $\mathbf{\Phi}\in\mathbb{C}^{N_\mathrm{R}\times N_\mathrm{R}}$ for the diagonal phase shifter.
In this case, the beamforming gain is given by
\begin{equation}
\beta=|\mathbf{g}^T\mathbf{\Phi}\mathbf{h}|^2.
\end{equation}
When the RIS perfectly aligns the phase of $\mathbf{g}$ and $\mathbf{h}$ (e.g., \cite{Yang2020,Basar2019,Abdullah2020,Tao2020}), the beamforming gain is given by
\begin{equation}\label{eqn72}
\beta=\Big(\sum_{n=0}^{N_\mathrm{R}-1}|g_n h_n|\Big)^2.
\end{equation}

However, it is challenging to directly analyze the outage probability based \eqref{eqn72}, as the PDF of $\beta$ is hardly available.
Hence, $\beta$ needs to be approximated for further analysis.
Fortunately, the CLT can serve as a suitable approximation for $\beta$, as outlined in \textit{Theorem~\ref{thm08}}.

\begin{thm}\label{thm08}
Assume 1) $N_\mathrm{R}\gg1$; 2) the RIS falls in the far field of the transmitter and receiver (i.e., $h_n g_n$ is i.i.d. $\forall n$).
Then, based on the CLT, $\sqrt{\beta}$ approximates a real-Gaussian RV with its mean and variance given by 
\begin{IEEEeqnarray}{rl}
\mathbb{E}(\sqrt{\beta})&\approx N_\mathrm{R}\mathbb{E}(|h_{n}g_{n}|),\\
\mathrm{Var}(\sqrt{\beta})&\approx N_\mathrm{R}\mathrm{Var}(|h_{n}g_{n}|).\label{eqn74}
\end{IEEEeqnarray}
Therefore, $\beta$ is approximately non-central $\chi^2$-distributed, with $K=1$, $\sigma^2=\mathrm{Var}(\sqrt{\beta})$, and
\begin{equation}\label{eq72Full}
\mathcal{M}^2=N_\mathrm{R}^2\mathbb{E}(|h_{n}g_{n}|)^2.
\end{equation}	
\end{thm}
\begin{IEEEproof}
The proof is straightforward and thus omitted. 
\end{IEEEproof}

\textit{Theorem~\ref{thm08}} indicates that \textbf{Algorithm~\ref{agthm2}} can also be employed to obtain the Cher-LB in RIS system.
Then, the transmission power can be adapted based on the Cher-LB to fulfill the single-shot outage probability as in \eqref{eqn53}.

Initially, the CLT was utilized to analyze $\beta$ in \cite{Basar2019} and subsequently employed in \cite{Yang2020, Abdullah2020}, where both $\mathbf{h}$ and $\mathbf{g}$ were assumed to be Rayleigh fading.
In \cite{Tao2020}, the authors extended this approximation to Rician fading scenarios. 
In this case, the mean and variance of $g_n h_n$ is given by
\begin{IEEEeqnarray}{rl}
\mathbb{E}(|g_n h_n|)&=\frac{\pi}{4}\mathcal{L}_{0.5}^{(0)}(-\kappa_h)\mathcal{L}_{0.5}^{(0)}(-\kappa_g),\label{eqn75}\\
\mathrm{Var}(|g_n h_n|)&=1-\frac{\pi^2(\mathcal{L}_{0.5}^{(0)}(-\kappa_h)\mathcal{L}_{0.5}^{(0)}(-\kappa_g))^2}{16(1+\kappa_h)(1+\kappa_g)}, 
\end{IEEEeqnarray}
where $\mathcal{L}(\cdot)$ stands for the generalized Laguerre function (see \cite{Tao2020}), $\kappa_h$ and $\kappa_g$ for the K-factor of $\mathbf{h}$ and $\mathbf{g}$, respectively.
The path loss of $\mathbf{h}$ and $\mathbf{g}$ is abbreviated, as it merely linearly scales the beamforming gain \cite{Tao2020}.

While the CLT has been extensively employed in RIS studies, there have been concerns regarding its inaccuracy (e.g.,\cite{Ding2020}). 
In our study, the inaccuracy may emerge when $N_\mathrm{R}$ is not significantly large (e.g., around $32$). 
Therefore, we examine the effectiveness and tightness of the Cher-LB through numerical results.

\subsection{Numerical Results and Discussion}

\begin{figure}[t]
\centering
\includegraphics[scale=0.53]{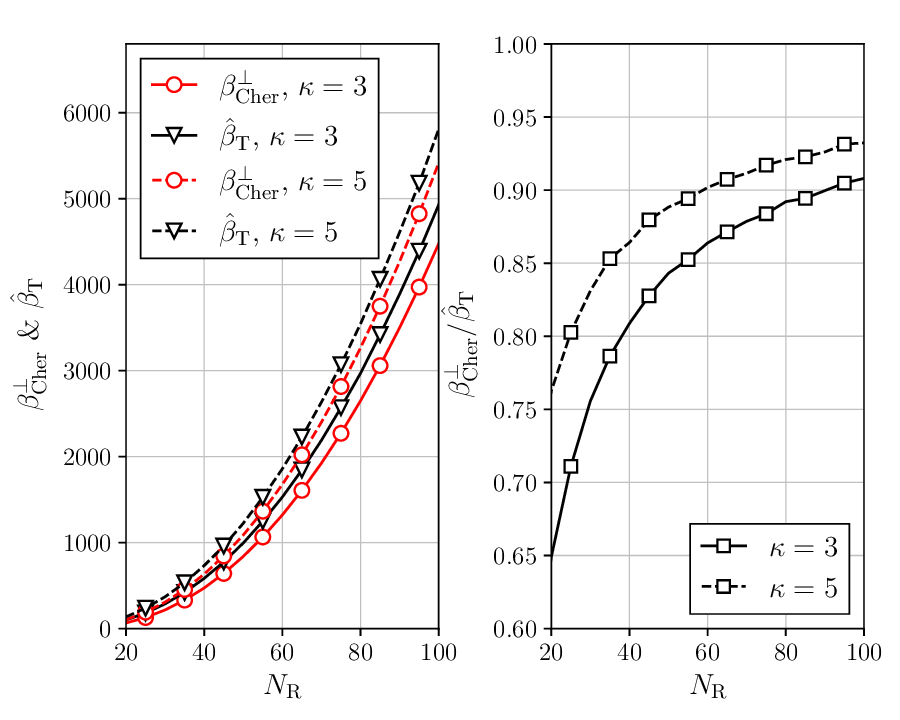}
\caption{The comparison between $\beta^\perp_\text{Cher}$ and $\hat{\beta}_{\mathrm{T}}$ in terms of beamforming gain as a function of $N_\mathrm{R}$ when $\kappa=3$ or $5$.}
\vspace{-1.3em}
\label{figRIS2}
\end{figure}

In this subsection, numerical results are employed to demonstrate the performance of the Cher-LB in RIS systems under Rician channel. 
Specifically, we consider single-stream transmission, where there is single-antenna at the transmitter and receiver (e.g., \cite{Basar2019,Tao2020}).
The outage requirement is considered to be $\epsilon=10^{-5}$ for the convenience to obtain $\hat{\beta}_\mathrm{T}$. 
With the decrease of $\epsilon$, the effectiveness of $\beta^\perp_{\text{Cher}}$ will remain the same.
For \textbf{Algorithm~\ref{agthm2}}, $\delta_{\beta}$ is still set to be $10^{-4}$.
It is assumed that the K-factor of the channel from the transmitter to the RIS and from the RIS to the receiver is the same (i.e., $\kappa_h=\kappa_g=\kappa$), where we consider $\kappa=3$ or $5$ to show the performance of the Cher-LB when LoS path is dominant for the channel.

{\em Numerical Example 1:}
The aim of this example is to examine the effectiveness of the Cher-LB when using CLT to approximate $\beta$.
Fig.~\ref{figRIS2} shows the comparison between $\beta^\perp_{\text{Cher}}$ and $\hat{\beta}_\mathrm{T}$ with the increase of $N_\mathrm{R}$, where the black lines stand for $\hat{\beta}_\mathrm{T}$ and the red lines for $\beta^\perp_{\text{Cher}}$.

In Fig.~\ref{figRIS2}(a), it is shown that $\beta^\perp_{\text{Cher}}$ is always smaller than $\hat{\beta}_\mathrm{T}$.
This proves that using the Cher-LB with CLT is effective to support URLLC systems with RIS.
Moreover, when $\kappa$ increases from $3$ to $5$, $\beta^\perp_{\text{Cher}}$ is closer to $\hat{\beta}_\mathrm{T}$.
This is because $\rho$ becomes higher with the increase of $\kappa$, and coincides to the observations in Fig.~\ref{fig1NonCentral}.
In Fig.~\ref{figRIS2}(b), the tightness of $\beta^\perp_{\text{Cher}}$ (i.e., $(\beta^\perp_{\text{Cher}})/(\hat{\beta}_\mathrm{T})$) is demonstrated.
It is shown that the ratio is increasing with the increase of $N_\mathrm{R}$.
This is reasonable, as the CLT is more accurate when $N_\mathrm{R}$ is large.
Moreover, Fig.~\ref{figRIS2}(b) explicitly delivers an important message that using the CLT will not violate the outage requirement even when $N_\mathrm{R}$ is not too large.
The is because the CLT is a pessimistic consideration for the tail distribution of $\beta$.
In other words, when $\beta$ is approximated by CLT, it is more likely to be close to zero or even negative. 

\begin{figure}[t]
\centering
\includegraphics[scale=0.53]{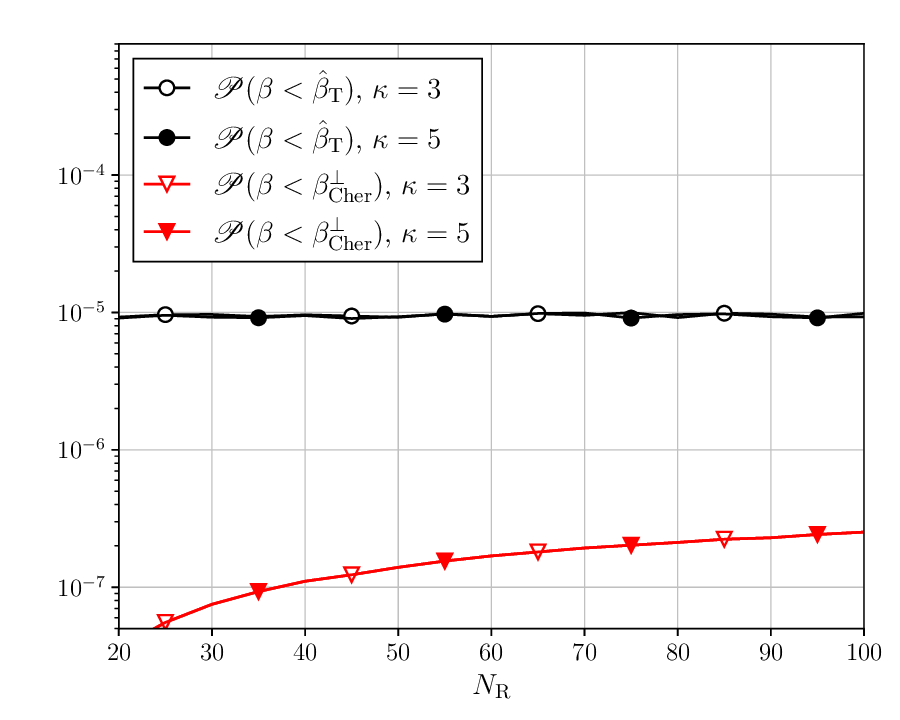}
\caption{The comparison between $\beta^\perp_\text{Cher}$ and $\hat{\beta}_{\mathrm{T}}$ in terms of outage probability as a function of $N_\mathrm{R}$ when $\kappa=3$ or $5$.}
\vspace{-1.3em}
\label{figRIS3}
\end{figure}

Fig.~\ref{figRIS3} demonstrates the outage probability of the Cher-LB compared to $\hat{\beta}_\mathrm{T}$.
It is observed that the outage probability of the Cher-LB is smaller than $\epsilon$ for two orders of magnitudes, which is even smaller than in Fig.~\ref{fig3NonCentral}.
This again shows that the CLT is a pessimistic consideration for the tail distribution of $\beta$.
With the increase of $N_\mathrm{R}$, the Cher-LB becomes closer to the $\hat{\beta}_\mathrm{T}$ as its outage probability is increasing.

\begin{figure}[t]
\centering
\includegraphics[scale=0.53]{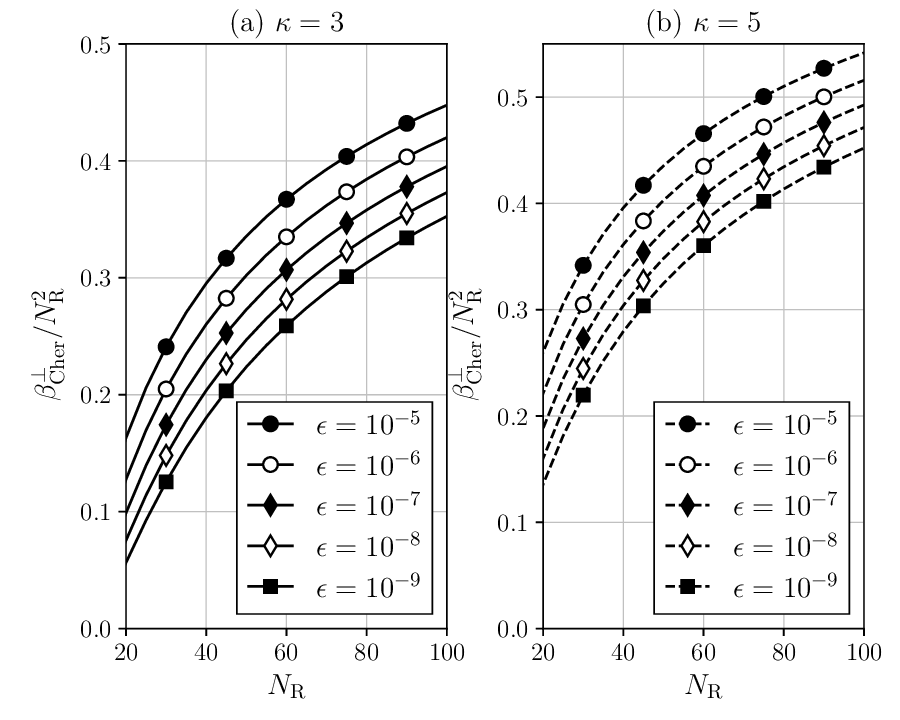}
\caption{$\beta^\perp_\text{Cher}$ normalized by $N_\mathrm{R}^2$ as a function of $N_\mathrm{R}$ with $\epsilon$ decreases from $10^{-5}$ to $10^{-9}$ when $\kappa=3$ or $5$.}
\vspace{-1.3em}
\label{figRIS1}
\end{figure}

{\em Numerical Example 2:}
The aim of this example is to demonstrate the improvement on the Cher-LB when the power gain of the RIS is eliminated.
The power gain of the RIS arises from concentrating the signal of different paths to the receiver.
\eqref{eqn75} shows that $\mathcal{M}^2\propto N_\mathrm{R}^2$, which means $\beta^\perp_{\text{Cher}}$ roughly quadratically with respect to $N_\mathrm{R}$, even when $\rho$ remains constant.
Therefore, we study the normalized beamforming gain (i.e., $(\beta^\perp_{\text{Cher}})/(N_\mathrm{R}^2)$) as shown in Fig.~\ref{figRIS1}.

It is observed that $(\beta^\perp_{\text{Cher}})/(N_\mathrm{R}^2)$ increases significantly with the increase of $N_\mathrm{R}$.
When $\kappa=3$, $(\beta^\perp_{\text{Cher}})/(N_\mathrm{R}^2)$ increases from around $0.1$ to around $0.4$, which accounts for around three-time improvement.
When $\kappa=5$, the increase is slightly less, i.e., $(\beta^\perp_{\text{Cher}})/(N_\mathrm{R}^2)$ increases from around $0.2$ to around $0.5$.
This is reasonable, as we have $\rho\propto N_\mathrm{R}$ by letting \eqref{eq72Full} divided by \eqref{eqn74}.
Moreover, as $\epsilon$ decreases to $10^{-9}$, the degradation of the normalized beamforming gain is not significant.
This coincides to \textit{Property~\ref{prpty03}}.

\section{Conclusion}\label{secIV}
In this paper, we have investigated approximations and lower bounds of the outage threshold when a RV obeys the non-central $\chi^2$ distribution.
It has been shown, through both theoretical and numerical analysis, that Cher-LB is the most effective lower bound amongst all studied candidates. 
The closed-form of Cher-LB was found hard to derive. 
Nevertheless, it could be obtained through line searching over a specific domain of the function. 
To develop a better understanding about the Cher-LB, we have rigorously established {three} of its properties with respect to the {mean, variance, reliability requirement, and degrees of freedom}. 
They are helpful to understand the behavior of the Cher-LB in communication systems.

The Cher-LB has been employed to form pessimistic prediction in communication systems both for MIMO-URLLC when the CSI-T obeys the first-order Markov process as well as for RIS systems where the beamforming gain is approximated by the CLT. 
It has been shown that the pessimistic prediction is made sufficiently accurate for the guaranteed reliability as well as the transmit-energy efficiency.

\appendices

\section{Proof of {Theorem \ref{thm05}}}\label{appdx2}
\begin{IEEEproof}
Given $\nu>0$, \eqref{eqn22}-\eqref{eqn23} show that $s(\nu, \beta^\perp)$ (or {$\psi(\nu, \beta^\perp)$}) increases monotonically with $\beta^\perp$. 
{Hence}, the minimum of $s(\nu,\beta^\perp)$ increases with $\beta^\perp$ and vice versa.

To find the interval where $\beta^\perp$ falls into, we are interested in the partial derivative of $s(\nu,\beta^\perp)$ to $\nu$, which yields
{\begin{IEEEeqnarray}{rl}
	&\frac{\partial s(\nu,\beta^\perp)}{\partial \nu}\IEEEnonumber\\
	=&s(\nu,\beta^\perp)\Bigl(\beta^\perp-\sum_{k=0}^{K-1}\Bigl(\frac{\mu_k^2}{(1+2\sigma_k^2\nu)^2}+\frac{\sigma_k^2}{1+2\sigma_k^2\nu}\Bigl)\Bigl)\IEEEnonumber\\
	\mathop{>}^\text{(c)}&s(\nu,\beta^\perp)\Bigl(\beta^\perp-\sum_{k=0}^{K-1}(\mu_k^2+\sigma_k^2)\Bigl).\label{eq24}
\end{IEEEeqnarray}}
The inequality (c) holds due to $(1+2\sigma_k^2\nu)>1$.
Then, we consider two cases of interval for $\beta^\perp$:

\textit{Case 1}: $\beta^\perp\in[\sum_{k=0}^{K-1}(\mu_k^2+\sigma_k^2), \infty)$. 
It is  immediately followed by: $\partial s(\nu,\beta^\perp)/\partial \nu>0$, and thus $s(\nu,\beta^\perp)$ is a monotonically increasing function of $\nu$. Then, the following result holds
\begin{equation}\label{eq25}
\inf_{\nu>0}s(\nu,\beta^\perp)=\lim_{\nu\to0}s(\nu,\beta^\perp)=1.
\end{equation}
Applying \eqref{eq25} into \eqref{eqn20} gives
\begin{equation}
\epsilon=\inf_{\nu>0}s(\nu,\beta^\perp)\geq 1.
\end{equation}
This result is not in line with the fact $\epsilon\in(0, 1)$, and thus \textit{Case 1} is not valid. 

\textit{Case 2}: $\beta^\perp\in(0, \sum_{k=0}^{K-1}(\mu_k^2+\sigma_k^2))$.
Let $\partial s(\nu,\beta^\perp)/\partial \nu=0$ and we obtain
\begin{equation}\label{eq26}
\beta^\perp-\sum_{k=0}^{K-1}\Bigl(\frac{\mu_k^2}{(1+2\sigma_k^2\nu)^2}+\frac{\sigma_k^2}{1+2\sigma_k^2\nu}\Bigl)=0.
\end{equation}
\eqref{eq26} is a polynomial equation with the order up to $2K$ (when $\sigma_0^2\neq\sigma_1^2\neq...\neq\sigma_{K-1}^2$). 
Hence, it would not be possible to obtain the closed-form solution in the general case. 
Fortunately, the summation in \eqref{eq26} decreases monotonically from $\sum_{k=0}^{K-1}(\mu_k^2+\sigma_k^2)$ to $0$ with the increase of $\nu$.
Therefore, there exists one and only one minimum of $s(\nu,\beta^\perp)$, and $\nu^\star$ that achieves this minimum can be obtained through line searching.

\eqref{eq26} also indicates
\begin{equation}\label{eq261}
\lim_{\nu^\star\rightarrow \infty}\beta^\perp=0.
\end{equation}
Moreover, we have 
{\begin{equation}\label{eq27}
	\lim\limits_{ \nu^\star\to\infty}\beta^\perp \nu^\star=\lim\limits_{\ \nu^\star\to\infty}\sum_{k=0}^{K-1}\Bigl(\frac{\mu_k^2\nu^\star+\sigma_k^2\nu^\star(1+2\sigma_k^2\nu^\star)}{(1+2\sigma_k^2\nu^\star)^2}\Bigl)=\frac{K}{2}.
\end{equation}}
By substituting \eqref{eq27} to \eqref{eqn22}, we obtain
\begin{equation}\label{eq271}
\lim_{\beta^\perp\rightarrow 0}s(\beta^\perp,\nu^\star)=0,
\end{equation}
or equivalently we have $\epsilon\rightarrow 0$ when $\beta^\perp\rightarrow 0$.
Moreover, when $\beta^\perp\rightarrow\sum_{k=0}^{K-1}(\mu_k^2+\sigma_k^2)$, we have $\nu^\star\to0$. 
In this case, \eqref{eqn22} gives
\begin{equation}\label{eq272}
\lim_{\nu^\star\rightarrow 0}s(\beta^\perp,\nu^\star)=1.
\end{equation}
In conclusion, as $\epsilon$ increases from $0$ to $1$, $\beta^\perp$ increases from $0$ to $\sum_{k=0}^{K-1}(\mu_k^2+\sigma_k^2)$. {\em Theorem~\ref{thm05}} is therefore proved. 
\end{IEEEproof}

\section{Proof of Property \ref{prpty01}}\label{appdxB}
\begin{IEEEproof}
Based on \eqref{eqn25}, the optimum $\tilde{\nu}$ is given by		
\begin{equation}\label{eq33}
\tilde{\nu}^\star=\frac{K\tilde{\sigma}^2+\sqrt{K^2\tilde{\sigma}^4+4\tilde{\beta}^\perp\widetilde{\mathcal{M}}^2}}{4\tilde{\sigma}^2\tilde{\beta}^\perp}-\frac{1}{2\tilde{\sigma}^2}. 
\end{equation}
It is known that $\tilde{\sigma}^2=\eta\sigma^2$ and $\widetilde{\mathcal{M}}^2=\eta\mathcal{M}^2$. 
Then, by substituting $\tilde{\beta}^\perp=\eta\beta^\perp$ to \eqref{eq33}, we have:
\begin{IEEEeqnarray}{rl}
1+2\tilde{\sigma}^2\tilde{\nu}^\star&=\frac{\eta K\sigma^2+\sqrt{\eta^2K^2\sigma^4+4\eta^2\beta^\perp\mathcal{M}^2}}{2\eta\beta^\perp}\label{eq36a}\\
&=\frac{K\sigma^2+\sqrt{K^2\sigma^4+4\beta^\perp\mathcal{M}^2}}{2\beta^\perp}\\
&=1+2\sigma^2\nu^\star.\label{eq36}
\end{IEEEeqnarray}
With \eqref{eq36a} and \eqref{eq36}, it is also easy to know that
\begin{equation}\label{eq37}
\eta\tilde{\nu}^\star=\nu^\star.
\end{equation}
By substituting \eqref{eq36} and \eqref{eq37} to \eqref{eqn24}, it can be computed that
\begin{equation}
s(\tilde{\nu}^\star,\eta\beta^\perp)=s(\nu^\star,\beta^\perp)=\epsilon.
\end{equation}
\textit{Property~\ref{prpty01}} is therefore proved.
\end{IEEEproof}

\section{Proof of Property \ref{prpty03}}\label{appdxB2}
\begin{IEEEproof}
It is mathematically hard to express $\lambda$ in $\varepsilon$ (or $\epsilon$).
Therefore, we translate \eqref{eqn32} into its equivalent form
\begin{equation}\label{eqn33a}
\lim\limits_{\lambda\to0}\frac{\partial\varepsilon}{\partial\lambda}=\infty.
\end{equation}
{
Applying \eqref{eqn16a} and \eqref{eqn16b} into \eqref{eqn24} with some tidy-up work yields
\begin{equation}\label{eqn29}
	\epsilon=\frac{\exp\left(\left(-\lambda(\rho+K)-\rho+\sqrt{K^2+4\lambda \rho(\rho+K)}\right)/2\right)}{\left((K+\sqrt{K^2+4\lambda \rho(\rho+K)})/(2\lambda (\rho+K))\right)^{\frac{K}{2}}}.
\end{equation}
We start by taking the natural logarithm on both sides of \eqref{eqn29} and obtain} 
\begin{IEEEeqnarray}{rl}
2\varepsilon\ln(10)&=-\lambda (\rho+K)-\rho+\sqrt{K^2+4\lambda \rho(\rho+K)}\IEEEnonumber\\
&~~~~+K\ln\Bigl(\frac{\sqrt{K^2+4\lambda \rho(\rho+K)}-K}{2\rho}\Bigl).\label{eqn34a}
\end{IEEEeqnarray}
Then, the partial derivative of $\varepsilon$ with respect to $\lambda$ is:
\begin{equation}\label{eqn35a}
\frac{2\ln(10)\partial\varepsilon}{\partial\lambda}=-(\rho+K)+\frac{K}{2\lambda}+\frac{2\rho(\rho+K)+\frac{K^2}{2\lambda}}{\sqrt{K^2+4\lambda\rho(\rho+K)}}.
\end{equation}
When $\lambda\to 0$, it is trivial to find \eqref{eqn33a} (or equivalently \eqref{eqn32}) true.

Moreover, the second-order partial derivative of $\varepsilon$ with respect to $\lambda$ is computed by
\begin{equation}\label{eqn36a}
\frac{2\ln(10)\partial^2\varepsilon}{\partial\lambda^2}=\frac{-K}{2\lambda^2}-\frac{\left(K^2+2\lambda\rho(\rho+K)\right)}{2\lambda^2\sqrt{K^2+4\lambda \rho(\rho+K)}}<0.
\end{equation}
It means $(\partial\varepsilon)/(\partial\lambda)$ monotonically decreases with respect to $\lambda$.
Hence, we have 
\begin{equation}\label{eq37a}
\frac{\partial\varepsilon}{\partial\lambda}>\lim\limits_{\lambda\to1}\frac{\partial\varepsilon}{\partial\lambda}\mathop{=}^\text{(d)}0,
\end{equation}
where the proof of (d) is a bit tedious but straightforward and thus omitted.
Given \eqref{eq37a}, the monotonicity of $\lambda$ with respect to $\epsilon$ is proved.
\end{IEEEproof}

\balance

\ifCLASSOPTIONcaptionsoff
\newpage
\fi

\bibliographystyle{myIEEEtran}
\bibliography{bibs/Bib_Else,bibs/Books_and_Standards,bibs/NLP_Downlink,bibs/GroupPaper,bibs/IterativePrecoding,bibs/digitalPreDistortion,bibs/URLLC}

\begin{IEEEbiography}[{\includegraphics[width=1in,height=1.25in,clip]{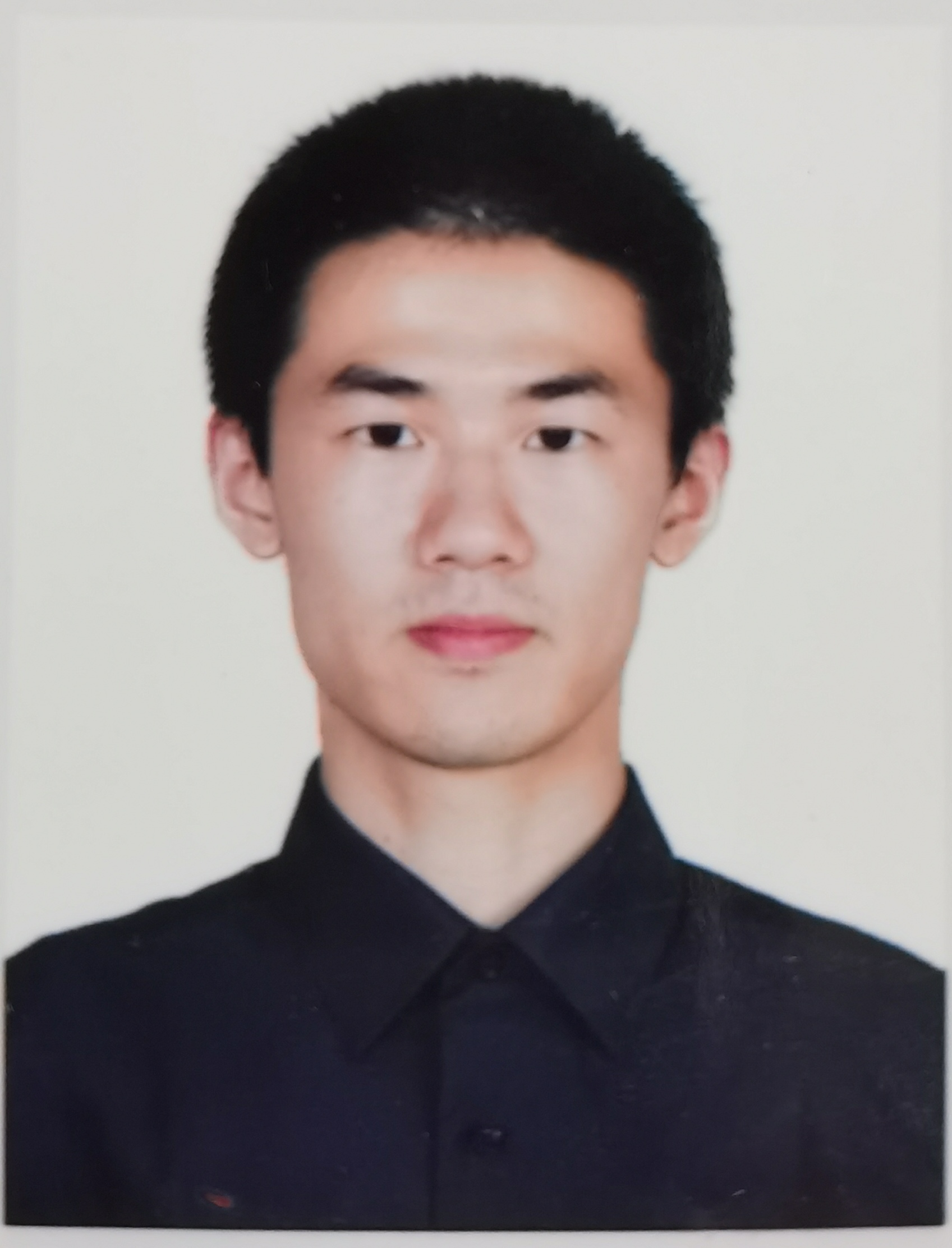}}]{Jinfei Wang}
received his Ph.D. degree from the University of Surrey, U.K., in 2023. He is currently a Research Fellow at the 5GIC\&6GIC, Institute for Communication Systems (ICS), University of Surrey, Guildford, U.K. His main research interests include: physical layer design of massive multiple-input multiple-output (MIMO) systems, ultra-reliable low-latency communication (URLLC), physical layer design of extremely large aperture array (ELAA) systems and stochastic process.
\end{IEEEbiography}

\begin{IEEEbiography}[{\includegraphics[width=1in,height=1.25in,clip]{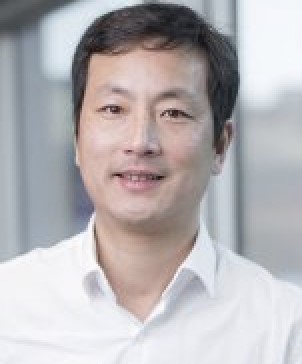}}]{Yi Ma} (Senior Member, IEEE) is a Chair Professor within the Institute for Communication Systems (ICS), University of Surrey, Guildford, U.K. He has authored or co-authored 200+ peer-reviewed IEEE journals and conference papers in the areas of deep learning, cooperative communications, cognitive radios, interference utilization, cooperative localization, radio resource allocation, multiple-input multiple-output, estimation, synchronization, and modulation and detection techniques. He holds 10 international patents in the areas of spectrum sensing and signal modulation and detection. He has served as the Tutorial Chair for EuroWireless2013, PIMRC2014, and CAMAD2015. He was the Founder of the Crowd-Net Workshop in conjunction with ICC’15, ICC’16, and ICC’17. He is the Co-Chair of the Signal Processing for Communications Symposium in ICC’19.
\end{IEEEbiography}

\begin{IEEEbiography}[{\includegraphics[width=1in,height=1.25in,clip]{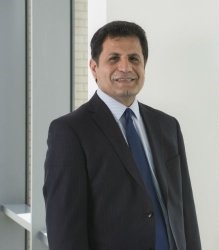}}]{Rahim Tafazolli}
(Senior Member, IEEE) is Regius Professor of Electronic Engineering, Professor of Mobile and Satellite Communications, Founder and Director of 5GIC, 6GIC and ICS (Institute for Communication System) at the University of Surrey. He has over 40 years of experience in digital communications research and teaching. He has authored and co-authored 1,000+ research publications and is regularly invited to deliver keynote talks and distinguished lectures to international conferences and workshops. He was the leader of study on “grand challenges in IoT” (Internet of Things) in the UK, 2011-2012, for RCUK (Research Council UK) and the UK TSB (Technology Strategy Board). He is the Editor of two books on Technologies for Wireless Future (Wiley) vol. 1, in 2004 and vol. 2, in 2006. He holds Fellowship of Royal Academy of Engineering (FREng), Institute of Engineering and Technology (FIET) as well as that of Wireless World Research Forum. He was also awarded the 28th KIA Laureate Award- 2015 for his contribution to communications technology.
\end{IEEEbiography}

\begin{IEEEbiography}[{\includegraphics[width=1in,height=1.25in,clip]{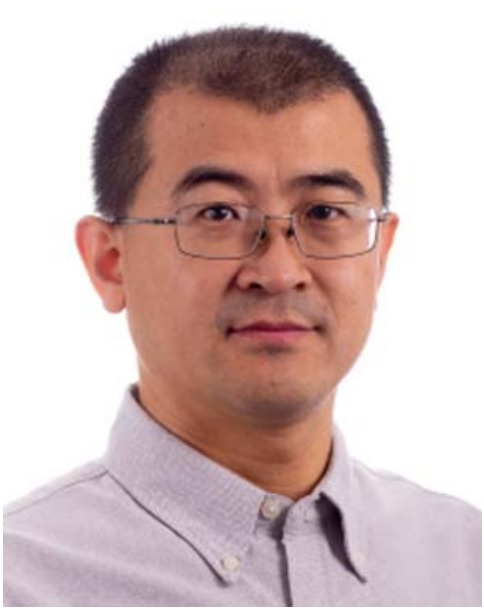}}]{Zhibo Pang}	
(Senior Member, IEEE) received the MBA degree in innovation and growth from the University of Turku, Turku, Finland, in 2012 and the Ph.D. in electronic and computer systems from the KTH Royal Institute of Technology, Stockholm, Sweden, in 2013. 

From 2019 to 2023, he was an Adjunct Professor with the University of Sydney, Camperdown, NSW, Australia. He is currently a Senior Principal Scientist with the ABB Corporate Research Sweden, Vasteras, Sweden, and an Adjunct Professor with the KTH Royal Institute of Technology. His research interests include enabling technologies in electronics, communication, computing, control, artificial intelligence, and robotics for Industry 4.0 and Healthcare 4.0.

Dr. Pang is a Member of the IEEE IES Industry Activities Committee, Steering Committee Member of the IEEE IoT Technical Community, Vice-Chair of the TC on Cloud and Wireless Systems for Industrial Applications, and Co-Chair of the TC on Industrial Informatics. He is an Associate Editor of IEEE TII, IEEE JBHI, and IEEE JESTIE. He was General Chair of IEEE ES2017, General Co-Chair of IEEE WFCS2021, and Invited Speaker at the Gordon Research Conference AHI2018. He was the recipient of the “Inventor of the Year Award” by ABB Corporate Research Sweden, three times in 2016, 2018, and 2021, respectively.
\end{IEEEbiography}

\end{document}